\documentclass{intech}

\booktitle{Will-be-set-by-IN-TECH}%

\chaptertitle{Quadratic discrete Fourier transform \\ and mutually unbiased bases} 

\authors{Maurice R. Kibler}
\affiliation{Universit\'e de Lyon, Universit\'e Claude Bernard et CNRS (IPNL et IN2P3)}
\country{France}

\begin{document}

\maketitle

\section{Introduction, motivations and notations}
The use of the discrete Fourier transform (DFT) is quite spread in many fields of physical sciences 
and engineering as for instance in signal theory. This chapter deals with a quadratic extension 
of the DFT and its application to quantum information. 

From a very general point of view, the DFT can be defined as follows. Let us denote 
($x_0$, $x_1$, $\ldots$, $x_{d-1}$) a collection of $d$ complex numbers. The transformation  
      \begin{eqnarray}
        x  \equiv (x_0, x_1, \ldots, x_{d-1}) \mapsto 
{\tilde x} \equiv ({\tilde x}_0, {\tilde x}_1, \ldots, {\tilde x}_{d-1})
      \label{definition DFT}
      \end{eqnarray}
defined by 
      \begin{eqnarray}
{\tilde x}_{\alpha}    =    \frac{1}{\sqrt{d}} \sum_{n = 0}^{d-1} e ^{i 2 \pi \alpha n / d} x_n, 
\quad \alpha = 0, 1, \ldots, d-1
      \label{DFT of numbers}
      \end{eqnarray}
will be referred to as the DFT of $x$. 

Equation (\ref{DFT of numbers}) can be transcribed in finite quantum mechanics. In that case, 
$x$ is often replaced by an orthonormal basis 
$\{ \vert n \rangle : n = 0, 1, \ldots, d-1 \}$ 
of the Hilbert space $\mathbb{C}^d$ (with an inner product noted $\langle \, \vert \, \rangle$ 
in Dirac notations). The analog of (\ref{DFT of numbers}) reads 
      \begin{eqnarray}
\vert \tilde{\alpha} \rangle     =    \frac{1}{\sqrt{d}} \sum_{n = 0}^{d-1} e ^{i 2 \pi \alpha n / d} \vert n \rangle, 
\quad \alpha = 0, 1, \ldots, d-1
      \label{quantum DFT}
      \end{eqnarray}
Equation (\ref{quantum DFT}) makes it possible to pass from the orthonormal basis 
$\{ \vert n \rangle : n = 0, 1, \ldots, d-1 \}$ 
to another orthonormal basis 
$\{ \vert \tilde{\alpha} \rangle : \alpha = 0, 1, \ldots, d-1 \}$
and vice versa since 
      \begin{eqnarray}
\langle n \vert n' \rangle = \delta (n,n') \Leftrightarrow  
\langle \tilde{\alpha} \vert \tilde{\alpha'} \rangle = \delta (\alpha,\alpha')
      \label{orthonormality}
      \end{eqnarray}
The transformation (\ref{quantum DFT}) defines a quantum DFT. In the last twenty years, 
the notion of quantum DFT has received a considerable attention in connection with finite 
quantum mechanics and quantum information \citep{Vourdas2004}. 

As an interesting property, the two bases 
$\{ \vert n \rangle              : n      = 0, 1, \ldots, d-1 \}$ and 
$\{ \vert \tilde{\alpha} \rangle : \alpha = 0, 1, \ldots, d-1 \}$, connected via a quantum DFT, 
constitute a couple of unbiased bases. Let us recall that two distinct orthonormal bases 
      \begin{eqnarray}
B_a    =    \{ | a \alpha \rangle : \alpha = 0, 1, \ldots, d-1 \}      
      \label{basis Ba}
      \end{eqnarray}
and 
      \begin{eqnarray}
B_b    =    \{ | b \beta  \rangle : \beta  = 0, 1, \ldots, d-1 \}   
      \label{basis Bb}
      \end{eqnarray}
of the space $\mathbb{C}^{d}$ are said to be unbiased if and only if 
      \begin{eqnarray}
\forall \alpha = 0, 1, \ldots, d-1, \ \ 
\forall \beta  = 0, 1, \ldots, d-1 \ : \ \vert \langle a \alpha | b \beta \rangle \vert = \frac{1}{\sqrt{d}}
      \label{definition of unbiased}
      \end{eqnarray} 
The unbiasedness character of the bases 
$\{ \vert n \rangle              : n      = 0, 1, \ldots, d-1 \}$ and 
$\{ \vert \tilde{\alpha} \rangle : \alpha = 0, 1, \ldots, d-1 \}$ then follows from 
      \begin{eqnarray}
      \langle n \vert \tilde{\alpha} \rangle       = \frac{1}{\sqrt{d}} e ^{i 2 \pi \alpha n / d} \ \ \Rightarrow  \ \ 
\vert \langle n \vert \tilde{\alpha} \rangle \vert = \frac{1}{\sqrt{d}}  
      \label{alpha et n unbiased}
      \end{eqnarray}
which is evident from (\ref{quantum DFT}). 

The determination of sets of mutually unbiased bases (MUBs) in $\mathbb{C}^{d}$ is of paramount 
importance in the theory of information and in quantum mechanics. Such bases are useful 
in classical information \citep{Calderbanketal1997}, quantum information 
\citep{Cerfetal2002} as well as for the construction of discrete Wigner 
functions \citep{Gibbonsetal2004}, the solution of the mean King problem \citep{EnglertAharonov2001} 
and the understanding of the Feynman path integral formalism \citep{Tolar2009}. It 
is well-known that the number $N_{\scriptscriptstyle MUB}$ of MUBs in $\mathbb{C}^d$ is such that 
$3 \leq N_{\scriptscriptstyle MUB} \leq d+1$ \citep{Durtetal2010}. Furthermore, the maximum number 
$N_{\scriptscriptstyle MUB} = d+1$ is reached when $d$ is a prime number or a power of a prime number 
\citep{Ivanovic1981,WoottersFields1989,Calderbanketal1997}. However, when $d$ is not a prime number 
or more generally a power of a prime number, it is not known if the limiting value $N_{\scriptscriptstyle MUB} = d+1$ 
is attained. In this respect, in the case $d=6$, in spite of an enormous number of works it was not possible to 
find more than three MUBs (see for example \citep{Grassl2005,Bengtssonetal2007,BrierleyWeigert2009}). 

The main aim of this chapter is to introduce and discuss a generalization of the DFTs defined by (\ref{DFT of numbers}) 
and (\ref{quantum DFT}) in order to produce other couples of MUBs. The generalization will be achieved by introducing 
quadratic terms in the exponentials in (\ref{DFT of numbers}) and (\ref{quantum DFT}) through the replacement of the 
linear term $\alpha n$ by a quadratic term $\xi n^2 + \eta n + \zeta$ with $\xi$, $\eta$ and $\zeta$ in 
$\mathbb{R}$. The resulting generalized DFT will be referred to as a quadratic DFT. 

The material presented in this chapter is organized in the following way. Section 2 is devoted to the study of 
those aspects of the representation theory of the group $SU(2)$ in a nonstandard basis which are of relevance 
for the introduction of the quadratic DFT. The quadratic DFT is studied in section 3. Some applications of the 
quadratic DFT to quantum information are given in section 4. 

Most of the notations in this chapter are standard. Some specific notations shall be introduced when 
necessary. As usual, $\delta_{a , b}$ stands for the Kronecker delta symbol of $a$ and $b$, $i$ for 
the pure imaginary, $\overline{z}$ for the complex conjugate of the number $z$, $A^{\dagger}$ for the 
adjoint of the operator $A$, and ${I}$ for the identity operator. We use $[A,B]_q$ to denote the 
$q$-commutator $AB - qAB$ of the operators $A$ and $B$; the commutator $[A,B]_{+1}$ and anticommutator 
$[A,B]_{-1}$ are noted simply $[A,B]$ and $\{ A,B \}$, respectively, as is usual in quantum 
mechanics. Boldface letters are reserved for squared matrices (${\bf I_d}$ is the $d$-dimensional
identity matrix). We employ a notation of type $\vert \psi \rangle$, or sometimes $\vert \psi )$, 
for a vector in an Hilbert space and we denote 
$\langle \phi \vert \psi \rangle$ and $\vert \phi \rangle \langle \psi \vert$ respectively the inner and 
outer products of the vectors $\vert \psi \rangle$ and $\vert \phi \rangle$. The symbols $\oplus$ and 
$\ominus$ refer to the addition and subtraction modulo $d$ or $2j+1$ (with $d = 2j+1 = 2, 3, 4, \ldots$ 
depending on the context) while the symbol $\otimes$ serves to denote the tensor product of two vectors 
or of two spaces. Finally $\mathbb{N}$, $\mathbb{N}^*$ and $\mathbb{Z}$ are the sets of integers, 
strictly positive integers and relative integers; $\mathbb{R}$ and $\mathbb{C}$ the real and complex 
fields; and $\mathbb{Z}/d\mathbb{Z}$ the ring of integers $0, 1, \ldots, d-1$ modulo $d$. 

\section{A nonstandard approach to $SU(2)$}
\subsection{Quon algebra}
The idea of a quon takes its origin in the replacement of the commutation (sign
$-$) and anticommutation (sign $+$) relations
 \begin{eqnarray}
 a_- a_+ \pm a_+ a_- = 1
 \end{eqnarray}
of quantum mechanics by the relation 
 \begin{eqnarray}
 a_- a_+ - q a_+ a_- = f(N)
 \end{eqnarray} 
where $q$ is a constant and $f(N)$ an arbitrary fonction of a number operator
$N$. The introduction of $q$ and $f(N)$ yields the possibility to replace the
harmonic oscillator algebra by a deformed oscillator algebra. 
For $f(N) = 1$, the case $q = -1$ corresponds to fermion operators (describing 
                a fermionic oscillator) and 
                the case $q = +1$             to boson   operators (describing
                a bosonic oscillator). The other possibilities for $q$ and
$f(N) = 1$ correspond to quon operators. We shall be concerned here with a quon 
algebra or $q$-deformed oscillator algebra for $q$ a root of unity. 

{\bf Definition 1}. {\it 
The three linear operators $a_-$, $a_+$ and $N_a$ such that
              \begin{eqnarray}
 [a_- , a_+]_q = I, \quad
 [ N_a , a_{+} ] = a_{+}, \quad
 [ N_a , a_{-} ] =-a_{-}, \quad
 \left( a_{+} \right)^k = \left( a_{-} \right)^k = 0, \quad 
 \left( N_a \right)^{\dagger} = N_a 
              \label{Aq(x)}
              \end{eqnarray}
where
              \begin{eqnarray}
q    =    \exp \left( \frac{2 \pi { i }}{k} \right), \quad k \in \mathbb{N} \setminus \{ 0,1 \}
              \label{def(q,k)}
              \end{eqnarray}
define a quon algebra or $q$-deformed oscillator algebra denoted  
$A_q(a_-, a_+, N_a)$ or simply $A_q(a)$. The operators $a_-$ and $a_+$ are referred to 
as quon operators. The operators $a_-$, $a_+$ and $N_a$ are called 
annihilation, creation and number operators, respectively.}
                       
Definition 1 differs from the one by Arik and Coon \citep{ArikCoon1976} in the sense that we 
take $q$ as a primitive $k$th root of unity instead of $0 < q < 1$. In Eq.~(\ref{def(q,k)}), 
the value $k = 0$ is excluded since it would lead to a non-defined 
value of $q$. The case $k = 1$ must be excluded too since it would yield trivial algebras
with $a_- = a_+ = 0$. We observe that for $k = 2$ (i.e., for $q = -1$), the algebra $A_{-1}(a)$ 
corresponds to the ordinary fermionic algebra and the quon operators coincide with the fermion 
operators. On the other hand, we note that in the limiting situation where $k \to \infty$ (i.e., 
for $q = 1$), the algebra $A_{1}(a)$ is nothing but the ordinary bosonic algebra and the quon 
operators are boson operators. For $k$ arbitrary, $N_a$ is generally different from $a_{+} a_{-}$; it 
is only for $k = 2$ and $k \to \infty$ that $N_a = a_{+} a_{-}$. Note that the nilpotency 
relations $\left( a_{+} \right)^k = \left( a_{-} \right)^k = 0$, with $k$ finite, are at the 
origin of $k$-dimensional representations of $A_{q}(a)$ (see section 2.2). 

For arbitrary $k$, the quon operators 
$a_-$ and $a_+$ are not connected via Hermitian conjugation. It is only for 
$k=2$ or $k \to \infty$ that we may take $a_+ = (a_-)^{\dagger}$. In general 
(i.e., for $k \not= 2$ or $k \not\to \infty$), we have 
$(a_{\pm})^{\dagger} \not= a_{\mp}$. Therefore, it is natural to consider the 
so-called $k$-fermionic algebra $\Sigma_q$ with the generators 
$a_-$, 
$a_+$, 
$a_+^+    =    (a_+)^{\dagger}$,  
$a_-^+    =    (a_-)^{\dagger}$ and $N_a$ \citep{Daoudetal1998}. The defining 
relations for $\Sigma_q$ correspond to the ones of 
$A_q       (a_-,   a_+,   N_a)$ and 
$A_{\bar q}(a_+^+, a_-^+, N_a)$ complemented by the relations
              \begin{eqnarray}                      
 a_- a_+^{+} - q^{- \frac{1}{2}} a_+^{+} a_- = 0, \quad 
 a_+ a_-^{+} - q^{  \frac{1}{2}} a_-^{+} a_+ = 0
              \end{eqnarray}                      
Observe that for $k = 2$ or $k \to \infty$, the latter relation 
corresponds to an identity. The operators $a_-$, $a_+$, $a_+^+$ and $a_-^+$ 
are called $k$-fermion operators and we also use the terminology $k$-fermions in analogy 
with fermions and bosons. They clearly interpolate between fermions and bosons. 

In passing, let us mention that the $k$-fermions introduced in \citep{Daoudetal1998}  
share some common properties with the parafermions of order $k-1$ discussed in 
\citep{RubakovSpiridonov1988,BeckersDebergh1990,Khare1993,Durand1993,KlishevichPlyushchay1999}. The 
$k$-fermions can be used for constructing a fractional supersymmetric algebra of order $k$ (or 
parafermionic algebra of order $k-1$). The reader may consult \citep{Daoudetal1998} for a 
study of the $k$-fermionic algebra $\Sigma_q$ and its application to supersymmetry. 

\subsection{Quon realization of $su(2)$}

Going back to quons, let us show how the Lie algebra $su(2)$ of the group $SU(2)$ can be 
generated from two quon algebras. We start with two commuting quon algebras $A_q(a)$  
with $a = x, y$ corresponding to the same value of the deformation parameter $q$. Their
generators satisfy Eqs.~(\ref{Aq(x)}) and (\ref{def(q,k)}) with $a = x, y$ and $[X , Y]=0$ 
for any $X$ in $A_q(x)$ and any $Y$ in $A_q(y)$. Then, let us look for Hilbertian representations 
of $A_q(x)$ and $A_q(y)$ on $k$-dimensional Hilbert spaces ${\cal F}_x$ and 
${\cal F}_y$ spanned by the bases   
$\{ | n_1 ) : n_1 = 0, 1, \ldots, k-1 \}$ and 
$\{ | n_2 ) : n_2 = 0, 1, \ldots, k-1 \}$, respectively. These two bases are supposed to be orthonormal, i.e., 
      \begin{eqnarray}                      
      ( n_1 |n_1' ) = \delta(n_1 , n_1'), \quad ( n_2 |n_2' ) = \delta(n_2 , n_2')                                            
      \end{eqnarray}   
We easily verify the following result. 

{\bf Proposition 1}. {\it 
The relations
      \begin{eqnarray}                      
  & & x_+ |n_1) = |n_1 + 1),                      \quad  x_+ |k-1) = 0   \nonumber \\                  
  & & x_- |n_1) = \left[ n_1   \right]_q |n_1-1), \quad  x_- |0)   = 0   \label{action des x} \\
  & & N_x |n_1) = n_1 |n_1)                                              \nonumber    
      \end{eqnarray}                      
and
      \begin{eqnarray}                      
& &  y_+ |n_2) = \left[ n_2+1 \right]_q |n_2+1),  \quad  y_+ |k-1) = 0 \nonumber \\
& &  y_- |n_2) = |n_2 - 1),                       \quad  y_- |0) = 0   \label{action des y} \\
& &  N_y |n_2) = n_2 |n_2)                                             \nonumber 
      \end{eqnarray}                      
define $k$-dimensional representations of $A_q(x)$ and $A_q(y)$, respectively. In (\ref{action des x}) and 
(\ref{action des y}), we use the notation 
      \begin{eqnarray}                      
\forall n \in \mathbb{N}^* \ : \ [n]_q    =    \frac{1-q^n}{1-q} = 1 + q + \ldots + q^{n-1}, \quad [0]_q    =    1
      \label{q-deformation of n}  
      \end{eqnarray}      
which is familiar in $q$-deformations of algebraic structures.}  
 
{\bf Definition 2}. {\it 
The cornerstone of the quonic approach to $su(2)$ is to define the two linear operators
              \begin{eqnarray}                        
  h    =    {\sqrt {N_x \left( N_y + 1 \right) }}, \quad v_{ra}    =    s_x s_y 
              \label{h and v_a n1}
              \end{eqnarray}                      
with  
\begin{eqnarray}
   s_x &   =   & q^{ a (N_x + N_y) / 2 } x_{+} + 
   {e} ^{ {i} \phi_r / 2 }  {1 \over 
  \left[ k-1 \right]_q!} (x_{-})^{k-1} 
  \label{esse x} \\
   s_y &   =   & y_{-}  q^{- a (N_x - N_y) / 2 }+ 
   {e} ^{ {i} \phi_r / 2 }  {1 \over 
  \left[ k-1 \right]_q!} (y_{+})^{k-1} 
  \label{esse y}
\end{eqnarray}
In (\ref{esse x}) and (\ref{esse y}), we take  
          \begin{eqnarray}                      
 a \in \mathbb{Z}/d\mathbb{Z}, \quad \phi_r = \pi (k-1) r, \quad r \in \mathbb{R} 
          \end{eqnarray}                      
and the $q$-deformed factorials are defined by 
     \begin{eqnarray}                      
\forall n \in \mathbb{N}^* \ : \ [n]_q !    =    [1]_q [2]_q \ldots [n]_q, \quad [0]_q !    =    1
      \label{q-factorial}  
      \end{eqnarray}
Note that the parameter $a$ might be taken as real. We limit ourselves to $a$ in $\mathbb{Z}/d\mathbb{Z}$ 
in view of the applications to MUBs.} 

The operators $h$ and $v_{ra}$ act on the states 
\begin{eqnarray}                      
| n_1 , n_2 ) = | n_1) \otimes | n_2 ) 
\end{eqnarray}                       
of the $k^2$-dimensional Fock space ${\cal F}_x \otimes {\cal F}_y$. It is straightforward 
to verify that the action of $v_{ra}$ on ${\cal F}_x \otimes {\cal F}_y$ is governed by
\begin{eqnarray} 
v_{ra} |k - 1, n_2) &=& e^{{i} \phi_r/2} |0, n_2 - 1), \quad n_2 \not= 0
\nonumber \\
v_{ra} |n_1 , n_2) &=& q^{n_2a} |n_1+1, n_2-1), \quad n_1 \not = k-1, \quad n_2 \not = 0
\label{vra sur n1n2} \\
v_{ra} |n_1 , 0) &=& e^{{i} \phi_r/2} |n_1 + 1, k -1), \quad n_1 \not= k-1
\nonumber
\end{eqnarray}
and
  \begin{eqnarray}
  v_{ra} |k-1, 0) = e^{{ i}{{\phi}_r }} |0, k-1)
  \end{eqnarray}
As a consequence, we can prove the identity
  \begin{eqnarray}
( v_{ra})^k = e^{{ i}{{\phi}_r }} I
  \end{eqnarray}
The action of $h$ on ${\cal F}_x \otimes {\cal F}_y$ is much more simple. It is described by
  \begin{eqnarray}
  h |n_1 , n_2) = \sqrt{ n_1 (n_2 + 1) } |n_1 , n_2)
  \end{eqnarray}
which holds for $n_1, n_2 = 0, 1, \ldots, k - 1$. Finally, the operator $v_{ra}$ is 
unitary and the operator $h$ Hermitian on the space ${\cal F}_x \otimes {\cal F}_y$.

We are now in a position to introduce a realization of the generators of the non-deformed 
Lie algebra $su(2)$ in terms of the operators $v_{ra}$ and $h$. As a preliminary step, let 
us adapt the trick used by Schwinger in his approach to angular momentum via a coupled pair 
of harmonic oscillators \citep{Schwinger1965}. This can be done by introducing two new 
quantum numbers $J$ and $M$ 
          \begin{eqnarray}                      
  J    =    {1 \over 2} \left( n_1+n_2 \right),  \quad  
  M    =    {1 \over 2} \left( n_1-n_2 \right) 
          \end{eqnarray} 
and the state vectors
          \begin{eqnarray}
  |J , M \rangle    =    |n_1 , n_2) = |J + M , J-M) \ \ \Rightarrow \ \ \langle J , M | J' , M' \rangle = \delta_{J , J'} \delta_{M , M'}
          \end{eqnarray}                        
Note that 
          \begin{eqnarray}                      
j    =    \frac{1}{2}(k-1)
          \end{eqnarray}                        
is an admissible value for $J$. We may thus have $j = {1 \over 2}, \> 1, \> {3 \over 2}, \> \ldots$  
(since $k = 2, \> 3, \> 4, \> \ldots$). For the value $j$ of $J$, the 
quantum number $M$ can take the values $m = j, j-1, \ldots, -j$. Then, let us consider the $(2j+1)$-dimensional
subspace $\epsilon(j)$ of the $k^2$-dimensional space 
${\cal F}_x \otimes {\cal F}_y$ spanned by the basis
          \begin{eqnarray}                      
B_{2j+1}    =    \{ |j , m \rangle : m = j, j-1, \ldots, -j \} 
          \label{standard SU(2) basis}
          \end{eqnarray} 
with the orthonormality property 
          \begin{eqnarray}                      
\langle j , m | j , m' \rangle = \delta_{m , m'}
          \end{eqnarray}                          
We guess that $\epsilon(j)$ is a space of constant angular momentum $j$. As a 
matter of fact, we can check that $\epsilon(j)$ is stable under $h$ and 
$v_{ra}$. 

{\bf Proposition 2}. {\it
The action of the operators $h$ and $v_{ra}$ on $\epsilon(j)$ 
is given by 
          \begin{eqnarray}                      
h      |j , m \rangle &=& \sqrt{(j+m)(j-m+1)} |j , m \rangle \label{action de h sur jm} \\
v_{ra} |j , m \rangle &=& \delta_{m,j} e^{{i} 2 \pi j r} |j , -j \rangle + (1 - \delta_{m,j}) q^{(j-m)a} |j , m+1 \rangle
          \label{action de vra sur jm}
          \end{eqnarray} 
where $q$ is given by (\ref{def(q,k)}) with $k=2j+1$, 
$r \in \mathbb{R}$ and $a \in \mathbb{Z}/(2j+1)\mathbb{Z}$.} 

It is sometimes useful to use the Dirac notation by writing
          \begin{eqnarray}
h &=& \sum_{m = -j}^{j} \sqrt{(j + m)(j - m+ 1)} 
\vert j, m \rangle \langle j, m \vert 
 \\
v_{ra} &=& {e}^{{i} 2 \pi j r} \vert j, -j \rangle \langle j, j \vert + \sum_{m=-j}^{j-1} q^{(j-m)a} \vert j, m+1 \rangle \langle j, m \vert 
\label{vra en notation de Dirac} \\
(v_{ra})^{\dagger} &=& {e}^{-{i} 2 \pi j r} \vert j, j \rangle \langle j,-j \vert + \sum_{m=-j+1}^{j} q^{-(j-m+1)a}\vert j, m-1 \rangle \langle j, m \vert 
          \end{eqnarray} 
It is understood that the three preceding relations are valid as far as the operators 
$h$, $v_{ra}$ and $(v_{ra})^{\dagger}$ act on the space $\epsilon(j)$. It is evident 
that $h$ is an Hermitian operator and $v_{ra}$ a unitary operator on $\epsilon(j)$.

{\bf Definition 3}. {\it 
The link with $su(2)$ can be established by introducing the three linear 
operators $j_{+}$, $j_{-}$ and $j_z$ through
              \begin{eqnarray}                      
  j_+    =    h           v_{ra},  \quad  
  j_-    =    \left( v_{ra} \right)^{\dagger} h,  \quad 
  j_z    =    \frac{1}{2} \left[ h^2 - \left( v_{ra} \right)^{\dagger} h^2 v_{ra} \right] 
              \label{definition of the j's}
              \end{eqnarray}                      
For each couple ($r,a$) we have a triplet ($j_+, j_-, j_z$). It is clear that $j_+$ and $j_-$ 
are connected via Hermitian conjugation and $j_z$ is Hermitian.} 

{\bf Proposition 3}. {\it 
The action of $j_{+}$, $j_{-}$ and $j_z$ on $\epsilon(j)$ is 
given by the eigenvalue equation
              \begin{eqnarray}                      
  j_z |j , m \rangle = m |j , m \rangle
              \label{jz sur m}   
              \end{eqnarray}
and the ladder equations
              \begin{eqnarray}   
  j_+ |j , m \rangle   & = & q^{ (j - m + s - 1/2)a}
  {\sqrt{ (j - m)(j + m+1) }} 
  |j , m + 1 \rangle 
              \label{jplus sur m} \\
  j_- |j , m \rangle   & = & q^{-(j - m + s + 1/2)a}
  {\sqrt{ (j + m)(j - m+1) }} 
  |j , m - 1 \rangle 
              \label{jmoins sur m}   
              \end{eqnarray}
where $s    =    1/2$.} 

For $a=0$, Eqs.~(\ref{jz sur m}), (\ref{jplus sur m}) and (\ref{jmoins sur m}) give relations that are 
well-known in angular momentum theory. Indeed, the case $a=0$ corresponds to the usual Condon and Shortley 
phase convention used in atomic and nuclear spectroscopy. As a corollary of Proposition 3, we have the 
following result. 

{\bf Corollary 1}. {\it 
The operators $j_+$, $j_-$ and $j_z$ satisfy the commutation 
relations
     \begin{eqnarray}
  \left[ j_z,j_{+} \right] =   j_{+},  \quad 
  \left[ j_z,j_{-} \right] = - j_{-},  \quad 
  \left[ j_+,j_- \right] = 2j_z 
     \label{adL su2}
     \end{eqnarray}
and thus span the Lie algebra of $SU(2)$.}

The latter result does not depend on the parameters $r$ and $a$. The writing of the ladder 
operators $j_+$ and $j_-$ in terms of $h$ and $v_{ra}$ constitutes a two-parameter polar 
decomposition of the Lie algebra $su(2)$. Thus, from two $q$-deformed oscillator algebras 
we obtained a polar decomposition of the non-deformed Lie algebra of $SU(2)$. This 
decomposition is an alternative to the polar decompositions obtained independently 
in \citep{Levy-Leblond1973,Vourdas1990,ChaichianEllinas1990}. 

\subsection{The $\{ j^2 , v_{ra} \}$ scheme}
Each vector $|j , m \rangle$ is a common eigenvector of the two commuting operators 
$j_z$ and      
     \begin{eqnarray}
     j^2    =    \frac{1}{2} \left( j_+j_- + j_-j_+ \right) + j_3^2 
          = j_+ j_- + j_3(j_3 - 1)
          = j_- j_+ + j_3(j_3 + 1) 
     \label{Casimir}
     \end{eqnarray}
which is known as the Casimir operator of $su(2)$ in group theory or as the square of a 
generalized angular momentum in angular momentum theory. More precisely, we have the eigenvalue 
equations
          \begin{eqnarray}
          j^2 |j , m \rangle = j(j+1) |j , m \rangle, \quad 
          j_z |j , m \rangle = m      |j , m \rangle, \quad m = j, j-1, \ldots, -j 
          \label{vp de j2 et jz}
          \end{eqnarray} 
which show that $j$ and $m$ can be interpreted as angular momentum quantum numbers (in units such 
that the rationalized Planck constant $\hbar$ is equal to $1$). Of course, the set $\{ j^2 , j_z \}$ 
is a complete set of commuting operators. It is clear that the two operators $j^2$ and $v_{ra}$ 
commute. As a matter of fact, the set $\{ j^2 , v_{ra} \}$ provides an alternative to the set 
$\{ j^2 , j_z \}$ as indicated by the next result. 

{\bf Theorem 1}. {\it 
For fixed $j$ (with $2j \in \mathbb{N}^*$), $r$ (with $r \in \mathbb{R}$) and $a$ 
(with $a \in \mathbb{Z}/(2j+1)\mathbb{Z}$), 
the $2j+1$ common eigenvectors of the operators $j^2$ and $v_{ra}$ can be taken in the form
              \begin{eqnarray}
|j \alpha ; r a \rangle = \frac{1}{\sqrt{2j+1}} \sum_{m = -j}^{j} 
q^{(j + m)(j - m + 1)a / 2 - j m r + (j + m)\alpha} | j , m \rangle, \quad
\alpha = 0, 1, \ldots, 2j 
              \label{j alpha r a in terms of jm}
              \end{eqnarray} 
where 
        \begin{eqnarray} 
        q    =    \exp \left( \frac{2 \pi { i }}{2j+1} \right)
        \label{definition of q en j}
        \end{eqnarray}
The corresponding eigenvalues are given by
          \begin{eqnarray}
          j^2 |j \alpha ; r a \rangle = j(j+1)              |j \alpha ; r a \rangle, \quad
       v_{ra} |j \alpha ; r a \rangle = q^{j(r+a) - \alpha} |j \alpha ; r a \rangle, \quad 
       \alpha = 0, 1, \ldots, 2j 
          \label{vp de j2 et vra}
          \end{eqnarray}           
so that the spectrum of $v_{ra}$ is nondegenerate and $\{ j^2 , v_{ra} \}$ does form a complete set of commuting 
operators. The inner product 
      \begin{eqnarray}
\langle j \alpha ; r a | j \beta ; r a \rangle = \delta_{\alpha,\beta}
      \label{jalphabetara}
      \end{eqnarray}
shows that 
      \begin{eqnarray}
B_{ra}    =    \{ |j \alpha ; r a \rangle : \alpha = 0, 1, \ldots, 2j \}
      \label{Bra basis}
      \end{eqnarray} 
is a nonstandard orthonormal basis for the irreducible matrix 
representation of $SU(2)$ associated with $j$. For fixed $j$, there 
exists {\it a priori} a ($2j+1$)-multiple infinity of orthonormal bases $B_{ra}$ since $r$ 
can have any real value and $a$, which belongs to the ring $\mathbb{Z}/(2j+1)\mathbb{Z}$, 
can take $2j+1$ values ($a = 0, 1, \ldots, 2j$).}

Equation (\ref{j alpha r a in terms of jm}) defines a unitary transformation that allows to pass 
from the standard orthonormal basis $B_{2j+1}$, quite well-known in angular momentum theory and group theory, 
to the nonstandard orthonormal basis $B_{ra}$. For
fixed $j$, $r$ and $a$, the inverse transformation of (\ref{j alpha r a in terms of jm}) is
      \begin{eqnarray}
\vert j,m \rangle = q^{-(j+m) (j-m+1) a/2 + jmr} \frac{1}{\sqrt{2j+1}}
\sum_{\alpha=0}^{2j} q^{-(j+m) \alpha} \vert j \alpha ; r a \rangle, \quad
m = j, j - 1, \ldots, -j
      \label{inverse of state jalphara}
      \end{eqnarray}
which looks like an inverse DFT up to phase factors. For $r=a=0$, Eqs.~(\ref{j alpha r a in terms of jm}) and 
(\ref{inverse of state jalphara}) lead to 
			\begin{eqnarray}
& & |j \alpha ; 0 0 \rangle = \frac{1}{\sqrt{2j+1}} \sum_{m = -j}^{j} 
q^{(j + m)\alpha} | j , m \rangle, \quad 
\alpha = 0, 1, \ldots, 2j 
			\label{j alpha 0 0 in terms of jm} \\
& \Leftrightarrow &
\vert j,m \rangle = \frac{1}{\sqrt{2j+1}}
\sum_{\alpha=0}^{2j} q^{-(j+m) \alpha} \vert j \alpha ; 0 0 \rangle, \quad
m = j, j - 1, \ldots, -j
			\label{inverse of state jalpha00}
			\end{eqnarray}
Equations (\ref{j alpha 0 0 in terms of jm}) and (\ref{inverse of state jalpha00}) correspond 
(up to phase factors) to the DFT of the basis $B_{2j+1}$ and its inverse DFT, respectively.

Note that the calculation of $\langle j \alpha ; r a | j \beta ; s b \rangle$ is much more involved for 
($r \not= s, \ a     = b$), 
($r     = s, \ a \not= b$) and 
($r \not= s, \ a \not= b$) than the one of $\langle j \alpha ; r a | j \beta ; r a \rangle$ (the value 
of which is given by (\ref{jalphabetara})). For example, the overlap between the
bases $B_{ra}$ and $B_{sa}$, of relevance for the case ($r \not= s, \ a     = b$), is given by
\begin{eqnarray}
\langle j \alpha ; ra | j \beta ; sa \rangle = \frac{1}{2j+1} \> 
\frac{\sin [j(s-r) + \alpha - \beta] \pi}
     {\sin [j(s-r) + \alpha - \beta] \frac{\pi}{2j+1}}
\end{eqnarray}
The cases ($r     = s, \ a \not= b$) and 
          ($r \not= s, \ a \not= b$) need the use of Gauss sums as we shall see below.

The representation theory and the Wigner-Racah algebra of the group $SU(2)$ 
can be developed in the $\{ j^2, v_{ra} \}$ quantization scheme. This 
leads to Clebsch-Gordan coefficients and $(3-j \alpha)_{ra}$ symbols with properties very different from the 
ones of the usual $SU(2) \supset U(1)$ Clebsch-Gordan coefficients and $3-jm$ symbols corresponding to 
the $\{ j^2, j_z \}$ quantization scheme. For more details, see Appendix which deals with the case $r=a=0$.       
         
The nonstandard approach to the Wigner-Racah algebra of $SU(2)$ and angular momentum theory in the 
$\{ j^2 , v_{ra} \}$ scheme is especially useful in quantum chemistry for problems involving cyclic symmetry. This 
is the case for a ring-shape molecule with $2j+1$ atoms at the vertices of a regular 
polygon with $2j+1$ sides or for a one-dimensional chain of $2j+1$ spins of $\frac{1}{2}$-value
each \citep{AlbKib2007}. In this connection, we observe that the vectors of type $|j \alpha ; r a \rangle$ 
are specific symmetry-adapted vectors. Symmetry-adapted vectors are widely used in quantum 
chemistry, molecular physics and condensed matter physics as for instance in ro-vibrational 
spectroscopy of molecules \citep{Championetal1977} and ligand-field theory \citep{Kibler1968}. However, 
the vectors $| j \alpha ; r a \rangle$ differ from the 
symmetry-adapted vectors considered in \citep{Kibler1968,PateraWinternitz1976,Championetal1977} 
in the sense that $v_{ra}$ is not an invariant under some finite subgroup (of crystallographic interest) 
of the orthogonal group $O(3)$. This can be clarified as follows. 

{\bf Proposition 4}. {\it From (\ref{vra en notation de Dirac}), it follows that the 
operator $v_{ra}$ is a pseudo-invariant under the cyclic group $C_{2j+1}$, a subgroup of $SO(3)$, 
whose elements are the Wigner operators $P_{R(\varphi)}$ associated with the rotations $R(\varphi)$, around 
the quantization axis $Oz$, with the angles
      \begin{eqnarray}
\varphi    =    p \frac{2 \pi}{2j+1}, \quad p = 0, 1, \ldots, 2j
      \end{eqnarray}
More precisely, $v_{ra}$ transforms as
      \begin{eqnarray}
P_{R(\varphi)} v_{ra} \left( P_{R(\varphi)} \right)^{\dagger} = e^{-i \varphi} v_{ra}
      \label{trans}
      \end{eqnarray}         
Thus, $v_{ra}$ belongs to the irreducible representation class of $C_{2j+1}$ 
of character vector 
      \begin{eqnarray}
\chi^{(2j)} = (1, q^{-1}, \ldots, q^{-2j})
      \end{eqnarray}
In terms of vectors of $\epsilon(j)$, we have 
      \begin{eqnarray}
P_{R(\varphi)} |j \alpha ; r a \rangle = q^{jp} |j \beta ; r a \rangle,  \quad \beta = \alpha \ominus p 
      \end{eqnarray}            
so that the set $\{ |j \alpha ; r a \rangle : \alpha = 0, 1, \ldots, 2j \}$ is stable under $P_{R(\varphi)}$. 
The latter set spans the regular representation of $C_{2j+1}$.}

\subsection{Examples}

{\bf Example 1}: The $j=\frac{1}{2}$ case. The eigenvectors of $v_{ra}$ are
     \begin{eqnarray}
\vert \frac{1}{2} \alpha ; ra \rangle = 
\frac{1}{\sqrt{2}} e^{{i} \pi(a/2 - r / 4 + \alpha)} \vert \frac{1}{2} , \frac{1}{2} \rangle + 
\frac{1}{\sqrt{2}} e^{{i} \pi r / 4} \vert \frac{1}{2} , -\frac{1}{2} \rangle, 
\quad  \alpha = 0, 1
     \label{eigenvectors of petitvra en 2 dim}
     \end{eqnarray}
where $r \in \mathbb{R}$ and $a$ can take the values $a = 0, 1$. In the case 
$r=0$, Eq.~(\ref{eigenvectors of petitvra en 2 dim}) gives the two bases
     \begin{eqnarray}
		B_{00} \ : \ \vert \frac{1}{2} 0 ; 00 \rangle =  \frac{1}{\sqrt{2}} \left(
  \vert \frac{1}{2} ,  \frac{1}{2} \rangle + \vert \frac{1}{2} , -\frac{1}{2} \rangle \right), \quad
                 \vert \frac{1}{2} 1 ; 00 \rangle = -\frac{1}{\sqrt{2}} \left(
  \vert \frac{1}{2} ,  \frac{1}{2} \rangle - \vert \frac{1}{2} , -\frac{1}{2} \rangle \right) 
     \label{dim2-1 en jm}
     \end{eqnarray}
and     
     \begin{eqnarray}
		B_{01} \ : \ \vert \frac{1}{2} 0 ; 01 \rangle = \frac{{i}}{\sqrt{2}} \left(
  \vert \frac{1}{2} ,  \frac{1}{2} \rangle - {i} \vert \frac{1}{2} , -\frac{1}{2} \rangle \right), \quad
                 \vert \frac{1}{2} 1 ; 01 \rangle = -\frac{i}{\sqrt{2}} \left(
  \vert \frac{1}{2} ,  \frac{1}{2} \rangle + {i} \vert \frac{1}{2} , -\frac{1}{2} \rangle \right) 
     \label{dim2-2 en jm}
     \end{eqnarray}       
The bases (\ref{dim2-1 en jm}) and (\ref{dim2-2 en jm}) 
are, up to phase factors, familiar bases 
in quantum mechanics for $\frac{1}{2}$-spin systems.

{\bf Example 2}: The $j=1$ case. The eigenvectors of $v_{ra}$ are
      \begin{eqnarray}
\vert 1 \alpha ; ra \rangle = \frac{1}{\sqrt{3}} q^{ r } 
\left( q^{a + 2 \alpha - 2 r} | 1, 1 \rangle + q^{a + \alpha - r} | 1, 0 \rangle + | 1, -1 \rangle \right), 
\quad \alpha = 0, 1, 2
      \label{eigenvectors of petitvra en 3 dim}
      \end{eqnarray}
where $r \in \mathbb{R}$ and $a$ can take the values $a = 0, 1, 2$. In the case 
$r=0$, Eq.~(\ref{eigenvectors of petitvra en 3 dim}) gives the three bases
     \begin{eqnarray}
B_{00} \ : \ \vert 1 0 ; 00 \rangle
& = & \frac{1  }{\sqrt{3}} \left(     | 1, -1 \rangle +     | 1, 0 \rangle +     | 1, 1 \rangle \right)  \nonumber  \\ 
             \vert 1 1 ; 00 \rangle 
& = & \frac{1  }{\sqrt{3}} \left(     | 1, -1 \rangle + q   | 1, 0 \rangle + q^2 | 1, 1 \rangle \right)  \label{62} \\ 
             \vert 1 2 ; 00 \rangle  
& = & \frac{1  }{\sqrt{3}} \left(     | 1, -1 \rangle + q^2 | 1, 0 \rangle + q   | 1, 1 \rangle \right)  \nonumber  
     \end{eqnarray}
     \begin{eqnarray}
B_{01} \ : \ \vert 1 0 ; 01 \rangle 
& = & \frac{1  }{\sqrt{3}} \left(     | 1, -1 \rangle + q    | 1, 0 \rangle + q  | 1, 1 \rangle \right)  \nonumber  \\ 
             \vert 1 1 ; 01 \rangle 
& = & \frac{1  }{\sqrt{3}} \left(     | 1, -1 \rangle + q^2 | 1, 0 \rangle +     | 1, 1 \rangle \right)  \label{63} \\
             \vert 1 2 ; 01 \rangle
& = & \frac{1  }{\sqrt{3}} \left(     | 1, -1 \rangle +     | 1, 0 \rangle + q^2 | 1, 1 \rangle \right)  \nonumber  
     \end{eqnarray}
     \begin{eqnarray}
B_{02} \ : \ \vert 1 0 ; 02 \rangle
& = & \frac{1  }{\sqrt{3}} \left(     | 1, -1 \rangle + q^2 | 1, 0 \rangle + q^2 | 1, 1 \rangle \right)  \nonumber  \\
             \vert 1 1 ; 02 \rangle 
& = & \frac{1  }{\sqrt{3}} \left(     | 1, -1 \rangle +     | 1, 0 \rangle + q   | 1, 1 \rangle \right)  \label{64} \\
             \vert 1 2 ; 02 \rangle 
& = & \frac{1  }{\sqrt{3}} \left(     | 1, -1 \rangle + q   | 1, 0 \rangle +     | 1, 1 \rangle \right)  \nonumber 
     \end{eqnarray}
It is worth noting that the vectors of the basis $B_{00}$ exhibit all characters 
     \begin{eqnarray}
\chi^{(\alpha)} = \left( 1, q^{\alpha}, q^{2 \alpha} \right), \quad \alpha = 0, 1, 2
     \label{characters of C3}
     \end{eqnarray} 
of the three vector representations of $C_3$. On another hand, the bases $B_{01}$ and $B_{02}$ 
are connected to projective representations of $C_3$ because they are described by the 
pseudo-characters
     \begin{eqnarray}
\chi_1^{(\alpha)} = \left( 1, q^{1 + \alpha}, q^{1 - \alpha} \right), \quad \alpha = 0, 1, 2
     \label{pseudo1characters of C3}
     \end{eqnarray} 
and
     \begin{eqnarray}
\chi_2^{(\alpha)} = \left( 1, q^{2 + \alpha}, q^{2 - \alpha} \right), \quad \alpha = 0, 1, 2
     \label{pseudo2characters of C3}
     \end{eqnarray}
respectively.
 
\section{Quadratic discrete Fourier transforms}

We discuss in this section two quadratic extensions of the DFT, namely, a 
quantum quadratic DFT that connects state vectors in a finite-dimensional 
Hilbert space, of relevance in quantum information, and a quadratic DFT 
that might be of interest in signal analysis. 

\subsection{Quantum quadratic discrete Fourier transform}
Relations of section 2 concerning $SU(2)$ can be transcribed in a form more adapted to 
the Fourier transformation formalism and to quantum information. In this respect, let 
us introduce the change of notations 
		\begin{eqnarray}
d    =    2j+1, \quad n    =    j+m, \quad | n \rangle    =    | j , -m \rangle
		\label{passage angular momentum QI1}
		\end{eqnarray}
and 
		\begin{eqnarray}
|a \alpha ; r \rangle    =    |j \alpha ; r a \rangle 
		\label{passage angular momentum QI2}
		\end{eqnarray}
so that (\ref{Bra basis}) becomes 
		\begin{eqnarray}
B_{ra} = \{ |a \alpha ; r \rangle : \alpha = 0, 1, \ldots, d-1 \}
		\label{new expression of Bra}
		\end{eqnarray}     
(Note that $d$ coincides with the dimension $k$ of the spaces ${\cal F}_x$ and ${\cal F}_y$ of section 1.) Then 
from Eq.~(\ref{j alpha r a in terms of jm}), we have  
      \begin{eqnarray}
\vert a \alpha ; r \rangle = 
q^{(d-1)^2 r / 4}
\frac{1}{\sqrt{d}} \sum_{n = 0}^{d-1}
q^{n(d -n) a/2 + n[\alpha -(d-1)r/2]} \vert d-1-n \rangle, \quad
\alpha = 0, 1, \ldots, d-1
      \label{aalphar en n}
      \end{eqnarray}
or equivalently
      \begin{eqnarray}
\vert a \alpha ; r \rangle = 
q^{(d-1)^2 r / 4}
\frac{1}{\sqrt{d}} \sum_{n = 0}^{d-1}
q^{(d-1-n)(n+1) a/2 + (d-1-n)[\alpha -(d-1)r/2]} \vert n \rangle, \quad 
\alpha = 0, 1, \ldots, d-1
      \label{aalphar en n bis}
      \end{eqnarray}      
where
				\begin{eqnarray} 
        q = \exp \left( \frac{2 \pi { i }}{d} \right)
        \label{definition of q en d}
        \end{eqnarray}
The inversion of (\ref{aalphar en n}) gives  
		    \begin{eqnarray} 
\vert d-1-n \rangle = q^{-n(d -n) a/2 - (d-1)^2 r/4 + n (d-1) r/2} \frac{1}{\sqrt{d}} 
\sum_{\alpha = 0}^{d-1} q^{-n \alpha} \vert a \alpha ; r \rangle, \quad
n = 0, 1, \ldots, d-1
				\label{inverse de aalphar}
        \end{eqnarray}
By introducing 
      \begin{eqnarray}
({\bf F_{ra}})_{n \alpha}    =   
\frac{1}{\sqrt{d}} q^{n(d -n) a/2 + (d-1)^2 r / 4 + n[\alpha -(d-1)r/2]}, \quad 
n, \alpha = 0, 1, \ldots, d-1
      \label{def Fra}
      \end{eqnarray}
equations (\ref{aalphar en n}) and (\ref{inverse de aalphar}) can be rewritten as 
      \begin{eqnarray}
\vert a \alpha ; r \rangle = \sum_{n = 0}^{d-1} \left( {\bf F_{ra}} \right)_{n \alpha} \, \vert d-1-n \rangle, 
\quad \alpha = 0, 1, \ldots, d-1
      \label{aalphar en n bis bis}
      \end{eqnarray}
and 
		    \begin{eqnarray} 
\vert d-1-n \rangle = \sum_{\alpha = 0}^{d-1} \overline{\left( {\bf F_{ra}} \right)_{n \alpha}} \, 
\vert a \alpha ; r \rangle, \quad n = 0, 1, \ldots, d-1
				\label{inverse de aalphar bis bis}
        \end{eqnarray}       
respectively. For $r=a=0$, Eqs.~(\ref{aalphar en n bis bis}) and (\ref{inverse de aalphar bis bis}) yield 
      \begin{eqnarray}
&& \vert 0 \alpha ; 0 \rangle = \frac{1}{\sqrt{d}} \sum_{n = 0}^{d-1}
e^{ {i} 2 \pi \alpha n / d} \vert d-1-n \rangle, \quad \alpha = 0, 1, \ldots, d-1 \nonumber \\
& \Leftrightarrow &
\vert d-1-n \rangle = \frac{1}{\sqrt{d}} \sum_{\alpha = 0}^{d-1}
e^{-{i} 2 \pi n \alpha / d} \vert 0 \alpha ; 0 \rangle, \quad n = 0, 1, \ldots, d-1
      \label{cas r et a nuls}
      \end{eqnarray}
which corresponds (up to a change of notations) to the DFT described by (\ref{quantum DFT}). For 
$a \not= 0$, Eq.~(\ref{aalphar en n bis bis}) can be considered as a quadratic extension (quadratic in $n$) 
of the DFT of the basis $\{ | n \rangle : n = 0, 1, \ldots, d-1 \}$ and Eq.~(\ref{inverse de aalphar bis bis}) 
thus appears as the corresponding inverse DFT. This can be summed up by the following definition. 

{\bf Definition 4}. {\it Let ${\bf H_{ra}}$ be the $d \times d$ matrix defined by the matrix elements 
      \begin{eqnarray}
({\bf H_{ra}})_{n \alpha}    =   
\frac{1}{\sqrt{d}} q^{(d-1-n)(n+1) a/2 + (d-1)^2 r / 4 + (d-1-n)[\alpha -(d-1)r/2]}, \quad 
n, \alpha = 0, 1, \ldots, d-1
      \label{def Hra}
      \end{eqnarray}
where, for a fixed value of $d$ (with $d \in \mathbb{N} \setminus \{ 0,1 \}$), $r$ and $a$ 
may have values in $\mathbb{R}$ and $\mathbb{Z}/d\mathbb{Z}$, respectively. In compact form 
      \begin{eqnarray}
({\bf H_{ra}})_{n \alpha} =
\frac{1}{\sqrt{d}} e^{2 \pi {i} \nu / d} 
      \label{def Fra en e1}
      \end{eqnarray}
with 
      \begin{eqnarray}
\nu    =   - \frac{1}{4} (d-1)^2 r 
           + \frac{1}{2} (d-1)a 
           + (d-1) \alpha 
           - \frac{1}{2} [ 2 \alpha + 2 a - d a - (d-1)r ] n 
           - \frac{1}{2} a n^2
      \label{def Fra en e2}
      \end{eqnarray}
The expansion 
      \begin{eqnarray}
\vert a \alpha ; r \rangle    =    \sum_{n = 0}^{d-1} \left( {\bf H_{ra}} \right)_{n \alpha} \, \vert n \rangle, 
      \quad
      \alpha = 0, 1, \ldots, d-1
      \label{def quadratic quantum DFT}
      \end{eqnarray}
defines a quadratic quantum DFT of the orthonormal basis 
		\begin{eqnarray}
B_d    =    \{ | n \rangle : n = 0, 1, \ldots, d-1 \}
    \label{base Bd}
		\end{eqnarray}
This transformation produces another orthonormal basis, namely, the basis $B_{ra}$ 
(see Eq.~(\ref{new expression of Bra})). The inverse transformation 
  	  \begin{eqnarray}
\vert n \rangle = \sum_{\alpha = 0}^{d-1} \overline{\left( {\bf H_{ra}} \right)_{n \alpha}} \, 
\vert a \alpha ; r \rangle, 
      \quad
      n = 0, 1, \ldots, d-1 
  	  \label{inverse of quadratic quantum DFT}
      \end{eqnarray}
gives back the basis $B_d$.}

For fixed $d$, $r$ and $a$, each of the $d$ vectors $\vert a \alpha ; r \rangle$, with $\alpha = 0, 1, \ldots, d-1$, 
is a linear combination of the vectors $\vert 0 \rangle, \vert 1 \rangle, \ldots, \vert d-1 \rangle$. The vector 
$\vert a \alpha ; r \rangle$ is an eigenvector of the operator
					\begin{eqnarray}
          v_{ra} = {e}^{{i} \pi (d-1) r} | d-1 \rangle \langle 0 | 
                  + \sum_{n=0}^{d-2} q^{(d-1-n)a} |d-2-n \rangle \langle d-1-n| 
					\label{definition of vra en d et n} 
          \end{eqnarray} 
or  
					\begin{eqnarray}
          v_{ra} = {e}^{{i} \pi (d-1) r} | d-1 \rangle \langle 0 | 
                  + \sum_{n=1}^{d-1} q^{na} |n-1 \rangle \langle n |     
          \label{definition of vra en d et n bis} 
          \end{eqnarray}		
(cf.~Eq.~(\ref{vra en notation de Dirac})). The operator $v_{ra}$ can be developed as
					\begin{eqnarray}
          v_{ra} = {e}^{{i} \pi (d-1) r} | d-1 \rangle \langle 0 | 
                 + q^{a}      | 0   \rangle \langle 1   |
                 + q^{2a}     | 1   \rangle \langle 2   |
                 + \ldots 
                 + q^{(d-1)a} | d-2 \rangle \langle d-1 |    
          \label{expansion of vra en d et n} 
          \end{eqnarray}	
Then, the action of $v_{ra}$ on the state $\vert n \rangle$ is described by
					\begin{eqnarray}
          v_{ra} \vert n \rangle = \delta_{n,0} e^{{i} \pi (d-1)r }\vert d-1 \rangle + (1 - \delta_{n,0}) q^{na} | n-1 \rangle 
          \label{action de vra sur n} 
          \end{eqnarray}
(cf.~Eq.~(\ref{action de vra sur jm})). Its eigenvalues are given by 
          \begin{eqnarray}
       v_{ra} |a \alpha ; r \rangle = q^{(d-1)(r+a)/2 - \alpha} |a \alpha ; r \rangle, \quad \alpha = 0, 1, \ldots, d-1 
          \label{vp de vra en d r a alpha}
          \end{eqnarray} 
(cf.~Eq.~(\ref{vp de j2 et vra})).

\subsubsection{Diagonalization of $v_{ra}$}
Let ${\bf V_{ra}}$ be the $d \times d$ 
unitary matrix that represents the linear operator $v_{ra}$ 
(given by (\ref{expansion of vra en d et n})) on the basis $B_d$. Explicitly, we have 
      \begin{eqnarray}
{\bf V_{ra}}    =   
		 \begin{pmatrix}
0                    &    q^a &      0  & \ldots &          0    \cr
0                    &      0 & q^{2a}  & \ldots &          0    \cr
\vdots               & \vdots & \vdots  & \ldots &     \vdots    \cr
0                    &      0 &      0  & \ldots & q^{(d-1)a}    \cr
e^{{i} \pi (d-1) r}  &      0 &      0  & \ldots &          0    \cr
     \end{pmatrix}       
     \label{matrix Vra}
     \end{eqnarray}
where the lines and columns are arranged in the order $0, 1, \ldots, d-1$. Note that 
the nonzero matrix elements of $V_{0a}$ are given by the irreducible character vector     
      \begin{eqnarray}
\chi^{(a)} = (1, q^a, \ldots, q^{(d-1)a})    
      \label{charaters of Cd and V0a}
      \end{eqnarray} 
of the cyclic group $C_d$.

{\bf Proposition 5}. {\it The matrix ${\bf H_{ra}}$ reduces the
endomorphism associated with the operator $v_{ra}$. In other words
      \begin{eqnarray}
\left( {\bf H_{ra}} \right)^{\dagger} {\bf V_{ra}} {\bf H_{ra}} =
q^{(d-1)(r+a) / 2} 
		 \begin{pmatrix}
q^0                  &      0 &       \ldots &          0    \cr
0                    & q^{-1} &       \ldots &          0    \cr
\vdots               & \vdots &       \ldots &     \vdots    \cr
0                    &      0 &       \ldots & q^{-(d-1)}    \cr
     \end{pmatrix}       
     \label{endomor}
     \end{eqnarray}
in agreement with Eq.~(\ref{vp de j2 et vra}).}

Concerning the matrices in (\ref{matrix Vra}) and (\ref{endomor}), it is important
to note the following convention. According to the tradition in quantum mechanics 
and quantum information, all the matrices in this chapter are set up with their 
lines and columns ordered from left to right and from top to bottom in the range 
$0, 1, \ldots, d-1$. Different conventions were used in some previous works by 
the author. However, the results previously obtained are equivalent to those 
of this chapter.

The eigenvectors of the matrix ${\bf V_{ra}}$ are 
      \begin{eqnarray}
\phi(a \alpha ; r) = \sum_{n = 0}^{d-1} \left( {\bf H_{ra}} \right)_{n \alpha}
\, \phi_{n}, \quad \alpha = 0, 1, \ldots, d-1
      \label{eigenvectors of Vra}
      \end{eqnarray}
where the $\phi_{n}$ with $n = 0, 1, \ldots, d-1$ are the column vectors
      \begin{eqnarray}	
\phi_0    =    
			\begin{pmatrix}
1      \cr
0      \cr
\vdots \cr
0      \cr
			\end{pmatrix}, \quad 
\phi_1    =   
			\begin{pmatrix}
0      \cr
1      \cr
\vdots \cr
0      \cr
			\end{pmatrix}, \quad 
\ldots, \quad 
\phi_{d-1}     =    
			\begin{pmatrix}
0      \cr
0      \cr
\vdots \cr
1      \cr
      \end{pmatrix}       
      \label{qudits en colonne}
      \end{eqnarray}
representing the state vectors $| 0 \rangle, | 1 \rangle, \ldots, | d-1 \rangle$, respectively. These 
eigenvectors are the column vectors of the matrix ${\bf H_{ra}}$. They satisfy the eigenvalue equation 
(cf.~\ref{vp de vra en d r a alpha})
      \begin{eqnarray}
{\bf V_{ra}} \phi(a \alpha ; r) = q^{(d-1)(r+a)/2 - \alpha} \phi(a \alpha ; r)
      \label{eq aux valeurs propres en matrices}
      \end{eqnarray}
with $\alpha = 0, 1, \ldots, d-1$. 

\subsubsection{Examples}

{\bf Example 3}: The $d=2$ case. For $d = 2$, there are two families of bases $B_{ra}$: 
the $B_{r0}$ family and the $B_{r1}$ family 
($a$ can take the values $a=0$ and $a=1$). In terms of matrices, we have 
      \begin{eqnarray}
{\bf H_{ra}}    =    \frac{1}{\sqrt{2}}
			\begin{pmatrix}
q^{  a/2 - r/4  }  &   -q^{  a/2 - r/4  } \cr
q^{  r/4  }        &    q^{  r/4  }       \cr
			\end{pmatrix}, \quad 
{{\bf V_{ra}}}    =   
			\begin{pmatrix}
0        &    q^a   \cr
q^{ r }  &      0   \cr
			\end{pmatrix}, \quad q = e^{i \pi} 
      \label{matrix Vra en 2 dim}
      \end{eqnarray}
The matrix ${\bf V_{ra}}$ has the eigenvectors (corresponding to the basis $B_{ra}$)
     \begin{eqnarray}
\phi ( a \alpha ; r ) = \frac{1}{\sqrt{2}} 
( q^{a/2 - r / 4 + \alpha} \phi_0 + q^{r / 4} \phi_1 ), \quad \alpha = 0, 1
     \label{eigenvectors of Vra en 2 dim}
     \end{eqnarray}
where 
      \begin{eqnarray}	
\phi_0    =    
			\begin{pmatrix}
1      \cr
0      \cr
			\end{pmatrix}, \quad 
\phi_1    =   
			\begin{pmatrix}
0      \cr
1      \cr
      \end{pmatrix}       
      \label{qubits en colonne}
      \end{eqnarray}
For $r=0$, we have
     \begin{eqnarray}
V_{00} = 
			\begin{pmatrix}
  0     &1   \cr
  1     &0   \cr
	  	\end{pmatrix}, \quad 
V_{01} = 
		  \begin{pmatrix}
  0     &-1  \cr
  1     &0   \cr
		  \end{pmatrix}
      \end{eqnarray}
the eigenvectors of which are (cf.~(\ref{eigenvectors of Vra en 2 dim})) 
     \begin{eqnarray}
						  \phi (0 0 ; 0) = \frac{1}{\sqrt{2}} \left(  \phi_1  +    \phi_0  \right)
						  =  \frac{1}{\sqrt{2}} 
		 \begin{pmatrix}
1      \cr
1      \cr
		 \end{pmatrix}, \quad
              \phi (0 1 ; 0) = \frac{1}{\sqrt{2}} \left(  \phi_1  -    \phi_0  \right)
              = -\frac{1}{\sqrt{2}} 
		 \begin{pmatrix}
1       \cr
-1      \cr
     \end{pmatrix}       
     \label{dim2-1}
     \end{eqnarray}
and     
     \begin{eqnarray}
							\phi (1 0 ; 0) = \frac{1}{\sqrt{2}} \left(  \phi_1  + i  \phi_0  \right)
							=  \frac{i}{\sqrt{2}} 
		 \begin{pmatrix}
1       \cr
-i      \cr
		 \end{pmatrix}, \quad
              \phi (1 1 ; 0) = \frac{1}{\sqrt{2}} \left(  \phi_1  - i  \phi_0  \right)
              = -\frac{i}{\sqrt{2}} 
     \begin{pmatrix}
1      \cr
i      \cr
     \end{pmatrix}       
     \label{dim2-2}        
     \end{eqnarray}
which correspond to the bases $B_{00}$ and $B_{01}$, respectively. Note that (\ref{dim2-1}) and 
(\ref{dim2-2}) are, up to unimportant multiplicative phase factors, qudits used in quantum information. 

{\bf Example 4}: The $d=3$ case. For $d=3$, we have three families of bases, 
that is to say $B_{r0}$, $B_{r1}$ and $B_{r2}$, 
since $a$ can be 0, 1 and 2. In this case 
      \begin{eqnarray}
{\bf H_{ra}}    =   \frac{1}{\sqrt{3}}
			\begin{pmatrix}
q^{a-r} & q^{a + 2 - r} & q^{a + 1 - r}  \cr
q^a     & q^{a + 1}     & q^{a + 2}      \cr
q^r     & q^r           & q^r            \cr
			\end{pmatrix}, \quad
{{\bf V_{ra}}}    =   
			\begin{pmatrix}
0             &    q^a &      0  \cr
0             &      0 & q^{2a}  \cr
q^{  3  r  }  &      0 &      0  \cr
			\end{pmatrix}, \quad q = e^{i 2 \pi / 3}
      \label{matrix Vra en dim 3}
      \end{eqnarray}
and ${\bf V_{ra}}$ admits the eigenvectors (corresponding to the basis $B_{ra}$)
      \begin{eqnarray}
\phi(a \alpha ; r) = \frac{1}{\sqrt{3}} q^{ r } 
\left( q^{a + 2 \alpha - 2 r} \phi_{0} + q^{a + \alpha - r} \phi_{1} + \phi_{2} \right), 
\quad \alpha = 0, 1, 2
      \label{eigenvectors of Vra en 3 dim}
      \end{eqnarray}
where
      \begin{eqnarray}	
\phi_0    =    
			\begin{pmatrix}
1      \cr
0      \cr
0      \cr
			\end{pmatrix}, \quad 
\phi_1    =   
			\begin{pmatrix}
0      \cr
1      \cr
0      \cr
			\end{pmatrix}, \quad 
\phi_{2}     =    
			\begin{pmatrix}
0      \cr
0      \cr
1      \cr
      \end{pmatrix}       
      \label{qutrits en colonne}
      \end{eqnarray}
In the case $r=0$, we get 
		\begin{eqnarray}
V_{00} = 
		\begin{pmatrix}
  0     &1   &0 \cr
  0     &0   &1 \cr
  1     &0   &0 \cr
		\end{pmatrix}, \quad 
V_{01} = 
		\begin{pmatrix}
  0     &q   &0   \cr
  0     &0   &q^2 \cr
  1     &0   &0   \cr
		\end{pmatrix}, \quad
V_{02} = 
		\begin{pmatrix}
  0     &q^2   &0 \cr
  0     &0     &q \cr
  1     &0     &0 \cr
		\end{pmatrix}
		\end{eqnarray}
The eigenvectors of $V_{00}$, $V_{01}$ and $V_{02}$ follow from  
Eq.~(\ref{eigenvectors of Vra en 3 dim}). This yields
     \begin{eqnarray}
         & & \phi(0 0 ; 0) = \frac{1}{\sqrt{3}} \left(     \phi_2 +     \phi_1 +     \phi_0 \right)  
         \nonumber  \\ 
         & & \phi(0 1 ; 0) = \frac{1}{\sqrt{3}} \left(     \phi_2 + q   \phi_1 + q^2 \phi_0 \right)  
         \label{010}\\ 
         & & \phi(0 2 ; 0) = \frac{1}{\sqrt{3}} \left(     \phi_2 + q^2 \phi_1 + q   \phi_0 \right)  
         \nonumber  
     \end{eqnarray}
or 
			\begin{eqnarray}
\phi(0 0 ; 0) = \frac{1}{\sqrt{3}} 
			\begin{pmatrix}
  1  \cr
  1  \cr
  1  \cr
			\end{pmatrix}, \quad
\phi(0 1 ; 0) = \frac{1}{\sqrt{3}} 
			\begin{pmatrix}
  q^2  \cr
  q    \cr
  1    \cr
			\end{pmatrix}, \quad
\phi(0 2 ; 0) = \frac{1}{\sqrt{3}} 
			\begin{pmatrix}
  q   \cr
  q^2 \cr
  1   \cr 
			\end{pmatrix}       
			\label{010col}
			\end{eqnarray}
corresponding to $B_{00}$, 
     \begin{eqnarray}
			   & & \phi(1 0 ; 0) = \frac{1}{\sqrt{3}} \left(     \phi_2 + q   \phi_1 + q   \phi_0 \right)  
			   \nonumber  \\ 
         & & \phi(1 1 ; 0) = \frac{1}{\sqrt{3}} \left(     \phi_2 + q^2 \phi_1 +     \phi_0 \right)  
         \label{110}\\
         & & \phi(1 2 ; 0) = \frac{1}{\sqrt{3}} \left(     \phi_2 +     \phi_1 + q^2 \phi_0 \right)  
         \nonumber  
     \end{eqnarray}
or 
			\begin{eqnarray}
\phi(1 0 ; 0) = \frac{1}{\sqrt{3}} 
			\begin{pmatrix}
  q  \cr
  q  \cr
  1  \cr
			\end{pmatrix}, \quad
\phi(1 1 ; 0) = \frac{1}{\sqrt{3}} 
			\begin{pmatrix}
  1    \cr
  q^2  \cr
  1    \cr
			\end{pmatrix}, \quad
\phi(1 2 ; 0) = \frac{1}{\sqrt{3}} 
			\begin{pmatrix}
  q^2    \cr
  1      \cr
  1      \cr 
			\end{pmatrix}       
			\label{110col}
			\end{eqnarray}     
corresponding to $B_{01}$, and 
     \begin{eqnarray}
				 & & \phi(2 0 ; 0) = \frac{1}{\sqrt{3}} \left(     \phi_2 + q^2 \phi_1 + q^2 \phi_0 \right)  
				 \nonumber  \\
         & & \phi(2 1 ; 0) = \frac{1}{\sqrt{3}} \left(     \phi_2 +     \phi_1 + q   \phi_0 \right)  
         \label{210}\\
         & & \phi(2 2 ; 0) = \frac{1}{\sqrt{3}} \left(     \phi_2 + q   \phi_1 +     \phi_0 \right)  
         \nonumber          
     \end{eqnarray}
or 		
			\begin{eqnarray}
\phi(2 0 ; 0) = \frac{1}{\sqrt{3}} 
			\begin{pmatrix}
  q^2  \cr
  q^2  \cr
  1    \cr
			\end{pmatrix}, \quad
\phi(2 1 ; 0) = \frac{1}{\sqrt{3}} 
			\begin{pmatrix}
  q  \cr
  1  \cr
  1  \cr
			\end{pmatrix}, \quad
\phi(2 2 ; 0) = \frac{1}{\sqrt{3}} 
			\begin{pmatrix}
  1  \cr
  q  \cr
  1  \cr
			\end{pmatrix}       
			\label{210col}
			\end{eqnarray}
corresponding to $B_{02}$. Note that (\ref{010col}), (\ref{110col}) and (\ref{210col}) are, 
up to unimportant multiplicative phase factors, qutrits used in quantum information.      

\subsubsection{Decomposition of $V_{ra}$} 
The matrix ${\bf V_{ra}}$ can be decomposed as
     \begin{eqnarray}
{\bf V_{ra}} = {\bf P_r} {\bf X} {\bf Z}^a
     \label{Vra en X et Z}
     \end{eqnarray}
where
       \begin{eqnarray}
{\bf P_r}    =   
			  \begin{pmatrix}
1                    &      0 &      0    & \ldots &       0                 \cr
0                    &      1 &      0    & \ldots &       0                 \cr
0                    &      0 &      1    & \ldots &       0                 \cr
\vdots               & \vdots & \vdots    & \ldots &  \vdots                 \cr
0                    &      0 &      0    & \ldots &     e^{{i} \pi (d-1) r} \cr
        \end{pmatrix}       
        \label{definition of P}
        \end{eqnarray}
and
       	\begin{eqnarray}
{\bf X}    =   
		  	\begin{pmatrix}
0                    &      1 &      0  & \ldots &       0 \cr
0                    &      0 &      1  & \ldots &       0 \cr
\vdots               & \vdots & \vdots  & \ldots &  \vdots \cr
0                    &      0 &      0  & \ldots &       1 \cr
1                    &      0 &      0  & \ldots &       0 \cr
				\end{pmatrix}, \quad
{\bf Z}    =   
			  \begin{pmatrix}
1                    &      0 &      0    & \ldots &       0       \cr
0                    &      q &      0    & \ldots &       0       \cr
0                    &      0 &      q^2  & \ldots &       0       \cr
\vdots               & \vdots & \vdots    & \ldots &  \vdots       \cr
0                    &      0 &      0    & \ldots &       q^{d-1} \cr
        \end{pmatrix}       
        \label{definition of X and Z}
        \end{eqnarray}
The matrices ${\bf X}$ and  ${\bf Z}$ can be derived from particular ${\bf V_{ra}}$ matrices since
    \begin{eqnarray}
{\bf X} = {\bf V_{00}}, \quad {\bf Z} = \left( {\bf V_{00}} \right)^{\dagger} {\bf V_{01}}
    \end{eqnarray}
which emphasize the important role played by the matrix ${\bf V_{ra}}$.

The matrices ${\bf P_r}$, ${\bf X}$ and ${\bf Z}$ (and thus ${\bf V_{ra}}$) are unitary. They satisfy
     \begin{eqnarray}
{\bf V_{ra}} {\bf Z} &=& q      {\bf Z} {\bf V_{ra}} \label{qcom 1} \\
{\bf V_{ra}} {\bf X} &=& q^{-a} {\bf X} {\bf V_{ra}} \label{qcom 2}
     \end{eqnarray}
Equation (\ref{qcom 1}) can be iterated to give the useful relation
     \begin{eqnarray}
({\bf V_{ra}})^m {\bf Z}^n = q^{mn}
{\bf Z}^n ({\bf V_{ra}})^m
     \label{VmZn}
     \end{eqnarray}
where $m,n \in \mathbb{Z}/d\mathbb{Z}$. Furthermore, we have the trivial relations 
     \begin{eqnarray}
e^{-{i} \pi (d-1) r} ({\bf V_{r0}})^d = {\bf Z}^d = {\bf I_d}     
     \label{nilpotency}
     \end{eqnarray}
More generally, we can show that
     \begin{eqnarray}
\forall n \in \mathbb{Z}/d\mathbb{Z} \ : \ ({\bf V_{ra}})^n = q^{-n(n-1)a/2} ({\bf V_{r0}})^n {\bf Z}^{an}
     \end{eqnarray}
Consequently
     \begin{eqnarray}
({\bf V_{ra}})^d = e^{{i} \pi (d-1) (r+a)} {\bf I_d}
     \end{eqnarray}
in agreement with the obtained eigenvalues for ${\bf V_{ra}}$ 
(see Eq.~(\ref{eq aux valeurs propres en matrices})).

\subsubsection{Weyl pairs} 
The relations in sections 3.1.1 and 3.1.3 can be particularized in the case $r = a = 0$. For 
example, Eq.~(\ref{VmZn}) gives the useful relation 
   \begin{eqnarray}
{\bf X}^m {\bf Z}^n = q^{mn}
{\bf Z}^n {\bf X}^m,
\quad  (m,n) \in \mathbb{N}^2      
   \label{XmZn}
   \end{eqnarray}
The fundamental relationship between the matrices ${\bf X}$ and ${\bf Z}$ is emphasized by the 
following proposition. 

{\bf Proposition 6}. {\it The unitary matrices ${\bf X}$ and ${\bf Z}$ satisfy the
$q$-commutation relation
     \begin{eqnarray}
[{\bf X} , {\bf Z}]_q = {\bf X} {\bf Z} - q {\bf Z} {\bf X} = 0      
     \label{XZ equal qZX}
     \end{eqnarray}
and the cyclicity relations
     \begin{eqnarray}
     {\bf X}^d =
     {\bf Z}^d = {\bf I_d}      
     \label{indempotency pour X et Z}
     \end{eqnarray}
In addition, they are connected through 
      \begin{eqnarray}
      \left( {\bf F_{00}} \right)^{\dagger} {\bf X} {\bf F_{00}} = {\bf Z}    
      \label{endomor pour X et Z}
      \end{eqnarray}
that indicates that ${\bf X}$ and ${\bf Z}$ are related by an ordinary DFT transform.}

According to Proposition 6, the matrices 
${\bf X}$ and ${\bf Z}$ constitute a Weyl pair (${\bf X}, {\bf Z}$). Weyl 
pairs were introduced at the beginning of quantum mechanics \citep{Weyl1931} 
and used for building operator unitary bases \citep{Schwinger1960}. We shall 
emphasis their interest for quantum information and quantum computing in 
section 4.

Let $x$ and $z$ be the linear operators associated with ${\bf X}$ and ${\bf Z}$, respectively. They are given by 
        \begin{eqnarray}
x = v_{00}, \quad z = (v_{00})^{\dagger} v_{01} \ \Rightarrow \ xz = v_{01}
    		\label{x z et xz} 
				\end{eqnarray}  
as functions of the operator $v_{ra}$. Each of the relations involving ${\bf X}$ and ${\bf Z}$ can be transcribed 
in terms of $x$ and $z$. 

The properties of $x$ follow from those of $v_{ra}$ with $r=a=0$. The unitary operator $x$ is 
a shift operator when acting on $\vert j , m \rangle$ or $\vert n \rangle$ (see (\ref{action de vra sur jm}) and 
(\ref{action de vra sur n})) and a phase operator when acting on 
$\vert j \alpha ; 00 \rangle = \vert 0 \alpha ; 0 \rangle$ 
(see (\ref{vp de j2 et vra}) and (\ref{vp de vra en d r a alpha})). More precisely, we have 
    \begin{eqnarray}
x | j,m \rangle = | j,m \oplus 1 \rangle \ \Leftrightarrow \ x | n \rangle = | n \ominus 1 \rangle 
    \label{x sur jm et sur n} 
	  \end{eqnarray}
and 
    \begin{eqnarray}
x | 0 \alpha ; 0 \rangle = q^{-\alpha} | 0 \alpha ; 0 \rangle
    \label{vp de x}  
	  \end{eqnarray}
The unitary operator $z$ satisfies
    \begin{eqnarray}
z | j,m \rangle = q^{j-m} | j,m \rangle 
\ \Leftrightarrow \ z | n \rangle = q^{n} | n \rangle
    \label{vp de z} 
	  \end{eqnarray}
and 
    \begin{eqnarray}
z | a \alpha ; 0 \rangle = q^{-1} | a \alpha_1 ; 0 \rangle, \quad \alpha_1 = \alpha \ominus 1 
    \label{z sur a alpha zero}
	  \end{eqnarray}
It thus behaves as a phase operator when acting on $\vert j , m \rangle$ or $\vert n \rangle$ and a shift 
operator when acting on $\vert a \alpha ; 0 \rangle$.  

In view of (\ref{vp de x}) and (\ref{vp de z}), the two cyclic operators $x$ and $z$ (cf.~$x^d = z^d = I$) 
are isospectral operators. They are connected via a discrete Fourier transform operator 
(see Eq.~(\ref{endomor pour X et Z})).

Let us now define the operators 
     \begin{eqnarray}
u_{ab} = x^a z^b, \quad a, b = 0, 1, \ldots, d-1 
     \end{eqnarray}
The $d^2$ operators $u_{ab}$ are unitary and satisfy the following trace relation 
          \begin{eqnarray}
 {\rm tr} \, \left( (u_{ab})^{\dagger} u_{a'b'} \right) = 
 d \>
 \delta_{a,a'} \> 
 \delta_{b,b'} 
          \label{trace de uu}
          \end{eqnarray}
where the trace is taken on the $d$-dimensional space $\epsilon(d) = \epsilon(2j+1)$. This trace relation 
shows that the $d^2$ operators $u_{ab}$ are pairwise orthogonal operators so that they 
can serve as a basis for developing any operator acting on the Hilbert space $\epsilon(d)$. Furthermore, 
the commutator and the anticommutator of $u_{ab}$ and $u_{a'b'}$ are given by 
          \begin{eqnarray}
[u_{ab} , u_{a'b'}] = \left( q^{-ba'} - q^{-ab'} \right) u_{a'' b''}, 
\quad a'' = a \oplus a', 
\quad b'' = b \oplus b' 
          \label{com new}
          \end{eqnarray}
and          
          \begin{eqnarray}
\{ u_{ab} , u_{a'b'} \} = \left( q^{-ba'} + q^{-ab'} \right) u_{a'' b''}, 
\quad a'' = a \oplus a',
\quad b'' = b \oplus b'
          \label{anticom new}
          \end{eqnarray}          
Consequently, $[u_{ab} , u_{a'b'}] = 0$ if and only if $ab' \ominus ba' = 0$ 
and $\{ u_{ab} , u_{a'b'} \} = 0$ if and only if $ab' \ominus ba' = (1/2) d$.  
Therefore, all anticommutators $\{ u_{ab} , u_{a'b'} \}$ are different from 0 
if $d$ is an odd integer. From a group-theoretical point of view, we have the 
following result. 

{\bf Proposition 7}. {\it The set $\{ u_{ab} = x^a z^b : a, b = 0, 1, \ldots, d-1 \}$ 
generates a $d^2$-dimensional Lie algebra. This algebra can be seen to be the Lie algebra of the general linear group  
$GL(d, \mathbb{C})$. The subset $\{ u_{ab} : a, b = 0, 1, \ldots, d-1 \} \setminus \{ u_{00} \}$ thus spans the Lie algebra of the special linear group $SL(d, \mathbb{C})$.} 

A second group-theoretical aspect connected with the operators $u_{ab}$ concerns a finite group, 
the so-called finite Heisenberg-Weyl group $WH(\mathbb{Z}/d\mathbb{Z})$, known as the Pauli group 
$P_d$ in quantum information \citep{Kibler2008}. The set $\{ u_{ab} : a, b = 0, 1, \ldots, d-1 \}$ is 
not closed under multiplication. However, it is possible to extend the latter set in order to 
have a group as follows. 

{\bf Proposition 8}. {\it Let us define the operators $w_{abc}$ via 
	\begin{eqnarray}
w_{abc} = q^a u_{bc}, \quad a, b, c = 0, 1, \ldots, d-1 
	\label{definition of wabc}
	\end{eqnarray}
Then, the set $\{ w_{abc} = q^a x^b z^c : a, b, c = 0, 1, \ldots, d-1 \}$, endowed with the 
multiplication of operators, is a group of order $d^3$ isomorphic with the Heisenberg-Weyl 
group $WH(\mathbb{Z}/d\mathbb{Z}$). This group, also referred to as the Pauli group $P_d$, 
is a nonabelian (for $d \geq 2$) nilpotent group with nilpotency class equal to 3. It is 
isomorphic with a finite subgroup of the group $U(d)$ for $d$ even or $SU(d)$ for $d$ odd.} 

Proposition 8 easily follows from the composition law
	\begin{eqnarray}
w_{abc} w_{a'b'c'} = w_{a''b''c''}, \quad 
a'' = a \oplus a' \ominus cb', \quad 
b'' = b \oplus b', \quad 
c'' = c \oplus c'  
	\label{composition law}
	\end{eqnarray}
Note that the group commutator of the two elements $w_{abc}$ and $w_{a'b'c'}$ of the group 
$WH(\mathbb{Z}/d\mathbb{Z})$ is
	\begin{eqnarray}
w_{abc} w_{a'b'c'} (w_{abc})^{-1} (w_{a'b'c'})^{-1} = w_{a''00}, \quad a'' = bc' \ominus cb' 
	\label{group commutator of the wabc}
	\end{eqnarray}
which can be particularized as
	\begin{eqnarray}
u_{ab} u_{a'b'} (u_{ab})^{-1} (u_{a'b'})^{-1} = q^{ab' \ominus ba'}I 
	\label{group commutator of the uab}
	\end{eqnarray}
in terms of the operators $u_{ab}$.

All this is reminiscent of the group $SU(2)$, the generators of which are the well-known Pauli matrices. 
Therefore, the operators $u_{ab}$ shall be referred as generalized Pauli operators and their matrices as 
generalized Pauli matrices. This will be considered further in section 4.

\subsubsection{Link with the cyclic group $C_d$} 
There exists an 
interesting connection between the operator $v_{ra}$ and the cyclic group
$C_d$ (see section 2). The following proposition presents another aspect of 
this connection.  

{\bf Proposition 9}. {\it Let $R$ be a generator of $C_d$ 
(e.g., a rotation of $2 \pi / d$ around an arbitrary axis). The application
       \begin{eqnarray}
R^n \mapsto {\bf X}^n \ : \ n = 0, 1, \ldots, d-1
       \end{eqnarray}
defines a $d$-dimensional matrix representation of $C_d$. This
representation is the regular representation of $C_d$.} 

Thus, the reduction of the representation $\{ {\bf X}^n : n = 0, 1, \ldots, d-1 \}$ 
contains once and only once each (one-dimensional) irreducible representation 
       \begin{eqnarray}
\chi^{(a)} = (1, q^{a}, \ldots, q^{(d-1)a}), \quad a = 0, 1, \ldots, d-1
       \end{eqnarray}
of $C_d$.

\subsubsection{Link with the $W_{\infty}$ algebra} 
Let us define the matrix
   \begin{eqnarray}
{\bf T_{(n_1,n_2)}}    =    q^{ {1 \over 2} n_1n_2 } {\bf Z}^{n_1} {\bf X}^{n_2}, \quad
                 (n_1,n_2) \in \mathbb{N}^2
   \label{ggggggg}
   \end{eqnarray}
It is convenient to use the abbreviation
   \begin{eqnarray}
(n_1, n_2) \equiv n  \ \ \Rightarrow  \ \ {\bf T_{(n_1,n_2)}} \equiv {\bf T_{n}}
   \label{92}
   \end{eqnarray}
The matrices ${\bf T_{n}}$ span an infinite-dimensional Lie algebra. This may be precised as follows.

{\bf Proposition 10}. {\it The commutator $[{\bf T_{m}},{\bf T_{n}}]$ is given by
   \begin{eqnarray}
[{\bf T_{m}} , {\bf T_{n}}] = - 2 {    i} \sin \left( { \pi \over d } m \times n \right) {\bf T_{m+n}} 
   \label{95}
   \end{eqnarray}
where
   \begin{eqnarray}
m \times n    =    m_1 n_2 - m_2 n_1, \quad m + n    =    (m_1 + n_1, m_2 + n_2)
   \label{94}
   \end{eqnarray}
The matrices ${\bf T_m}$ can be thus formally viewed as the generators of the infinite-dimensional 
Lie algebra $W_{\infty}$.} 

The proof of (\ref{95}) is easily obtained by using (\ref{XmZn}). This leads to 
   \begin{eqnarray}
{\bf T_m} {\bf T_n} = q^{ - {1 \over 2} m \times n } {\bf T_{m + n}}
   \label{93}
   \end{eqnarray}
which implies (\ref{95}). Thus, we get of Lie algebra $W_{\infty}$ (or sine algebra) 
investigated in \citep{Fairlieetal1990}. 

\subsection{Quadratic discrete Fourier transform}
\subsubsection{Generalities} 
We are now prepared for discussing analogs of the transformations (\ref{def quadratic quantum DFT}) 
and (\ref{inverse of quadratic quantum DFT}) in the language of classical signal theory. 

{\bf Definition 5}. {\it Let us consider the transformation 
    	\begin{eqnarray}
x = \{ x_{m    }  \in {\mathbb{C}} : m = 0, 1, \ldots, d-1 \}  \ \ \leftrightarrow  \ \
y = \{ y_{n    }  \in {\mathbb{C}} : n = 0, 1, \ldots, d-1 \}
    	\label{transformation x - y}
      \end{eqnarray}
defined by
    \begin{eqnarray}
y_{n} = \sum_{m = 0}^{d-1}           \left( {\bf F_{ra}} \right)_{m n}  \, x_{m} \ \ \Leftrightarrow \ \
x_{m} = \sum_{n = 0}^{d-1} \overline{\left( {\bf F_{ra}} \right)_{m n}} \, y_{n}
    \label{transfo yx et xy en Fra}
    \end{eqnarray}
where
      \begin{eqnarray}
({\bf F_{ra}})_{n m}    =   
\frac{1}{\sqrt{d}} q^{n(d -n) a/2 + (d-1)^2 r / 4 + n[m -(d-1)r/2]}, 
\quad n, m = 0, 1, \ldots, d-1
      \label{def Fra bis}
      \end{eqnarray}
For $a \not= 0$, the bijective transformation $x \leftrightarrow y$ can
be thought of as a quadratic DFT.} 

In Eq.~(\ref{transfo yx et xy en Fra}), we choose the matrix ${\bf F_{ra}}$ as the quadratic Fourier 
matrix instead of the matrix ${\bf H_{ra}}$ because the particular case $r=a=0$ corresponds to the 
ordinary DFT (see also \citep{AKW2010}). Note that the matrices ${\bf F_{ra}}$ and 
${\bf H_{ra}}$ are interrelated via
      \begin{eqnarray}
({\bf F_{ra}})_{n m}    =   ({\bf H_{ra}})_{n' m}, \quad n' = d-1-n
      \label{relation Fra Hra}
      \end{eqnarray}
Therefore, the lines of ${\bf F_{ra}}$ in the order $0, 1, \ldots, d-1$ coincide with those of 
${\bf H_{ra}}$ in the reverse order $d-1, d-2, \ldots, 0$.

The analog of the
Parseval-Plancherel theorem for the ordinary DFT can be expressed in
the following way. 

{\bf Theorem 2}. {\it The quadratic transformations $x \leftrightarrow
y$ and $x' \leftrightarrow y'$ associated with the same matrix ${\bf
F_{ra}}$, with $r \in \mathbb{R}$ and $a \in \mathbb{Z}/d\mathbb{Z}$,
satisfy the conservation rule
    \begin{eqnarray}
\sum_{n = 0}^{d-1} \overline{y_{n}} \, y'_{n} =
\sum_{m = 0}^{d-1} \overline{x_{m}} \, x'_{m}
    \label{Parseval-Plancherel}
      \end{eqnarray}
where both sums do not depend on $r$ and $a$.}  

\subsubsection{Properties of the quadratic DFT matrix} 
In order to get familiar with the quadratic DFT defined by (\ref{transfo yx et xy en Fra}), 
we now examine some of the properties of the quadratic DFT matrix ${\bf F_{ra}}$. 

{\bf Proposition 11}. {\it For $d$ arbitrary, the matrix elements of ${\bf F_{ra}}$ satisfies the 
useful symmetry properties 
   		 \begin{eqnarray}
&& \left( {\bf F_{ra}} \right)_{d-1 \alpha} = 
q^{(d-1)(r+a)/2 - \alpha} e^{ -i \pi (d-1) r } \left( {\bf F_{ra}} \right)_{0 \alpha}, 
\quad \alpha = 0, 1, \ldots, d-1 \\
&& \left( {\bf F_{ra}} \right)_{n-1 \alpha} = 
q^{(d-1)(r+a)/2 - \alpha + na} \left( {\bf F_{ra}} \right)_{n \alpha}, 
\ n = 1, 2, \ldots, d-1, \ \alpha = 0, 1, \ldots, d-1 
 		   \label{symmetry properties of Fra}
       \end{eqnarray}
which can be reduced to the sole symmetry relation 
   		 \begin{eqnarray}
 \left( {\bf F_{0a}} \right)_{n \ominus 1 \alpha} = 
 q^{(d-1)a/2 - \alpha + na} \left( {\bf F_{0a}} \right)_{n \alpha},                  
\quad n, \alpha = 0, 1, \ldots, d-1
 		   \label{symmetry properties of F0a}
       \end{eqnarray}
when $r=0$.}

{\bf Proposition 12}. {\it For $d$ arbitrary, the matrix ${\bf F_{ra}}$ is unitary.}

The latter result can be checked from a straightforward calculation. It also follows in a simple way from 
   		 \begin{eqnarray}
\langle j \alpha ; ra \vert j \beta ; sb \rangle 
= \langle a \alpha ; r  \vert b \beta ; s  \rangle = ( ({\bf F_{ra}})^{\dagger} {\bf F_{sb}} )_{\alpha \beta}
 		   \label{matrix Fra and inner product}
       \end{eqnarray}
It is sufficient to put $s=r$ and $b=a$ in (\ref{matrix Fra and inner product}) and to use (\ref{jalphabetara}).

For $d$ arbitrary, in addition to be unitary the matrix ${\bf F_{ra}}$ is such that the
modulus of each of its matrix elements is equal to $1/{\sqrt{d}}$. Thus, ${\bf F_{ra}}$ 
can be considered as a generalized Hadamard matrix (we adopt here the normalization of 
Hadamard matrices generally used in quantum information and quantum computing 
\citep{Kibler2009}). In the case where $d$ is a prime number, 
we shall prove in section 4 from (\ref{matrix Fra and inner product}) that the matrix 
$({\bf F_{ra}})^{\dagger} {\bf F_{rb}}$ is another Hadamard matrix for $b \not= a$. Similar 
results hold for the matrix ${\bf H_{ra}}$.

{\bf Proposition 13}. {\it For $d$ arbitrary, the matrix ${\bf F_{ra}}$ can be factorized as
      \begin{eqnarray}
{\bf F_{ra}} = {\bf D_{ra}} {\bf F}, \quad {\bf F}    =    {\bf F_{00}}
   	  \label{factorization of Fra}
      \end{eqnarray}
where ${\bf D_{ra}}$ is the $d \times d$ diagonal matrix with the matrix elements
      \begin{eqnarray}
({\bf D_{ra}})_{m n}    =    q^{m(d -m) a/2 + (d-1)^2 r / 4 - m(d-1)r/2} \delta_{m , n}
      \label{matrix D}
      \end{eqnarray}
and ${\bf F}$ is the well-known ordinary DFT matrix.}

For fixed $d$, there is one $d$-multiple infinity of Gaussian
matrices ${\bf D_{ra}}$ (and thus ${\bf F_{ra}}$) distinguished by
$a \in \mathbb{Z}/d\mathbb{Z}$ and $r \in \mathbb{R}$. The matrix
${\bf F}$ was the object of a great number of studies. The main
properties of the ordinary DFT matrix ${\bf F}$ are summarized in 
\citep{AKW2010}. Let us simply recall here the fundamental 
property 
      \begin{eqnarray}
{\bf F}^4 = {\bf I_d}
      \label{matrix F4}
      \end{eqnarray}
of interest for obtaining the eigenvalues and eigenvectors of ${\bf F}$.

{\bf Proposition 14}. {\it The determinant of ${\bf F_{ra}}$ reads 
      \begin{eqnarray}
{\rm det} \, {\bf F_{ra}} = e^{{i} \pi (d^2-1) a / 6} {\rm det} \, {\bf F}
      \label{determinant of Fra}
      \end{eqnarray}
where the value of ${\rm det} \, {\bf F}$ is well-known \citep{Mehta1987,AKW2010}.}

{\bf Proposition 15}. {\it The trace of ${\bf F_{ra}}$ reads
    \begin{eqnarray}
{\rm tr} \, {\bf F_{ra}} = e^{{i} \pi (d-1)^2 / (2d)} \frac{1}{\sqrt{d}} S(u, v, w)
    \label{trace of Fra}
      \end{eqnarray}
where $S(u, v, w)$ is
 			\begin{eqnarray}
S(u, v, w)    =    \sum_{k = 0}^{|w|-1} e^{{i} \pi (u k^2 + v k) / w}
      \label{Gauss sum}
      \end{eqnarray} 
with
    \begin{eqnarray}
u    =    2-a, \quad v    =    d(a-r) + r, \quad w    =    d
    \label{x-y-z}
    \end{eqnarray}
(note that $v$ is not necessarily an integer).}

Let us recall that the sum defined by (\ref{Gauss sum})
is a generalized quadratic Gauss sum. It can be calculated easily in the situation 
where $u$, $v$ and $w$ are integers such that $u$ and $w$ are mutually prime, $u w$ 
is not zero, and $uw + v$ is even \citep{Berndtetal1998}.

Note that the case $a = 2$ deserves a special attention. In this
case, the quadratic character of ${\rm tr} \, {\bf F_{ra}}$
disappears. In addition, if $r = 0$ we get
    \begin{eqnarray}
{\rm tr} \, {\bf F_{02}} = \sqrt{d}
    \label{trace of F02}
      \end{eqnarray}
as can be seen from a direct calculation.

{\bf Example 5}: In order to illustrate the preceding properties, let us consider the 
matrix 
      \begin{eqnarray}
      {\bf F_{02}} = \frac{1}{\sqrt{6}} 
      \begin{pmatrix}
1   &   1     &   1   &   1    &   1    &   1  \cr
q^5 &   1     &   q   &   q^2  &   q^3  & q^4  \cr
q^2 &   q^4   &   1   &   q^2  &   q^4  &   1  \cr
q^3 &   1     &   q^3 &   1    &   q^3  &   1  \cr
q^2 &   1     &   q^4 &   q^2  &   1    & q^4  \cr
q^5 &   q^4   &   q^3 &   q^2  &   q    &   1  \cr
      \end{pmatrix}       
      \label{d equal 6}
      \end{eqnarray}
corresponding to $d=6$ ($\Rightarrow q = e^{i \pi / 3}$), $r=0$ and $a=2$.  It is a simple matter of trivial 
calculation to check that the properties given above for ${\bf F_{ra}}$ are satisfied by the matrix ${\bf F_{02}}$.

\section{Application to quantum information}

	\subsection{Computational basis and standard $SU(2)$ basis} 

In quantum information science, we use qubits which are indeed state vectors in the Hilbert space 
$\mathbb{C}^2$. The more general qubit 
		\begin{eqnarray}
| \psi_2 \rangle    =    c_0 | 0 \rangle + c_1 | 1 \rangle, 
\quad c_0 \in \mathbb{C}, 
\quad c_1  \in \mathbb{C}, 
\quad |c_0|^2 + |c_1|^2 = 1 
		\label{psideux}
		\end{eqnarray} 
is a linear combination of the vectors $| 0 \rangle$ and $| 1 \rangle$ which constitute an orthonormal basis
		\begin{eqnarray}
B_2    =    \{ | 0 \rangle, | 1 \rangle \}
		\label{the B2 basis}
		\end{eqnarray}
of $\mathbb{C}^2$. These two vectors can be considered as the basis vectors for the fundamental irreducible 
representation class of $SU(2)$, in the $SU(2) \supset U(1)$ scheme, corresponding to $j=1/2$ with
\begin{eqnarray}
| 0 \rangle \equiv | 1/2,  1/2 \rangle, \quad | 1 \rangle \equiv | 1/2, -1/2 \rangle
\end{eqnarray} 
More generally, in $d$ dimensions we use qudits of the form
\begin{eqnarray}
| \psi_d \rangle    =    \sum_{n = 0}^{d-1} c_n | n \rangle,                                                      
\quad c_n \in \mathbb{C}, 
\quad n = 0, 1, \ldots, d-1, 
\quad \sum_{n = 0}^{d-1} |c_n|^2 = 1 
\end{eqnarray} 
where the vectors $| 0 \rangle, | 1 \rangle, \ldots, | d-1 \rangle$ span an 
orthonormal basis of $\mathbb{C}^d$ with 
		\begin{eqnarray}
\langle n | n' \rangle  = \delta_{n , n'}
		\label{orthonormality in n}
		\end{eqnarray}
By introducing
		\begin{eqnarray}
j    =    \frac{1}{2} (d-1), \quad m    =    n - \frac{1}{2} (d-1), \quad | j,m \rangle    =    | d-1-n \rangle 
		\label{passage QI angular momentum}
		\end{eqnarray}
(a change of notations equivalent to (\ref{passage angular momentum QI1})), 
the qudits $| n \rangle$ can be viewed as the basis vectors $| j,m \rangle$ 
for the irreducible representation class associated with $j$ of $SU(2)$ in the 
$SU(2) \supset U(1)$ scheme. More precisely, the correspondence between 
angular momentum states and qudits is 
		\begin{eqnarray}
  | 0   \rangle \equiv | j , j   \rangle, \quad
  | 1   \rangle \equiv | j , j-1 \rangle, \quad \ldots, \quad
  | d-1 \rangle \equiv | j , -j  \rangle
		\end{eqnarray}
where $| j , j   \rangle, | j , j-1 \rangle, \ldots, | j , -j  \rangle$ are common 
eigenvectors of angular momentum operators $j^2$ and $j_z$. In other words, 
the basis $B_d$ (see (\ref{base Bd})), known in quantum information 
as the computational basis, may be identified to the $SU(2) \supset U(1)$ standard 
basis or angular momentum basis $B_{2j+1}$ (see (\ref{standard SU(2) basis})). We 
shall see in section 4.2 that such an identification is very useful when $d$ is a 
prime number and does not seem to be very interesting when $d$ is not a prime integer. Note 
that the qudits $| 0   \rangle, | 1   \rangle, \ldots, | d-1 \rangle$ are often represented 
by the column vectors $\phi_0, \phi_1, \ldots, \phi_{d-1}$ (given by (\ref{qudits en colonne})), 
respectively. 

	\subsection{Mutually unbiased bases}

The basis $B_{ra}$ given by (\ref{new expression of Bra}) can serve as another basis for 
qudits. For arbitrary $d$, the couple ($B_{ra}, B_d$) exhibits an interesting property. For fixed 
$d$, $r$ and $a$, Eq.~(\ref{aalphar en n}) gives 
      \begin{eqnarray}
\forall n \in \mathbb{Z}/d\mathbb{Z}, \ 
\forall \alpha \in \mathbb{Z}/d\mathbb{Z} \ : \ 
\vert \langle n | a \alpha ; r \rangle \vert = \frac{1}{\sqrt{d}} 
      \label{Bra et Bd}
      \end{eqnarray}
Equation (\ref{Bra et Bd}) shows that $B_{ra}$ and $B_d$ are two unbiased bases.  

Other examples of unbiased bases can be obtained for $d = 2$ and $3$. We easily verify that the bases $B_{r0}$ and 
$B_{r1}$ for $d=2$ given by (\ref{eigenvectors of petitvra en 2 dim}) are unbiased. Similarly, the bases $B_{r0}$, $B_{r1}$ 
and $B_{r2}$ for $d=3$ given by (\ref{eigenvectors of petitvra en 3 dim}) are mutually unbiased. Therefore, by combining 
these particular results with the general result implied by (\ref{Bra et Bd}) we end up with three 
MUBs for $d=2$ and four MUBs for $d=3$, in agreement with $N_{\scriptscriptstyle MUB} = d+1$ 
when $d$ is a prime number. The results for $d=2$ and $3$ can be generalized 
in the case where $d$ is a prime number. This leads to the following theorem 
\citep{KibPla2006,AlbKib2007,Kibler2008}.

{\bf Theorem 3}. {\it For $d=p$, with $p$ a prime number, the bases $B_{r0}$, $B_{r1}$, $\ldots$, $B_{rp-1}$, 
$B_{p}$ corresponding to a fixed value of $r$ form a complete set of $p+1$ MUBs. The $p^2$ vectors 
$| a \alpha ; r \rangle$ or $\phi(a \alpha ; r)$, with $a, \alpha = 0, 1, \ldots, p-1$, of the bases  
$B_{r0}, B_{r1}, \ldots, B_{rp-1}$ are given by a single formula, namely, Eq.~(\ref{aalphar en n bis}) 
or (\ref{eigenvectors of Vra}). The index $r$ makes it possible to distinguish different complete 
sets of $p+1$ MUBs.}

The proof is as follows. First, according to (\ref{Bra et Bd}), the computational basis $B_{p}$ is 
unbiased with any of the $p$ bases $B_{r0}, B_{r1}, \ldots, B_{rp-1}$. Second, we get 
    \begin{eqnarray}	                  
\langle a \alpha ; r | b \beta ; r \rangle = \frac{1}{p}     
\sum_{k = 0}^{p-1} q^{k(p-k)(b-a) / 2 + k(\beta - \alpha)} 
    \label{produit scalaire} 
	  \end{eqnarray} 
or 
     \begin{eqnarray}
\langle a \alpha ; r | b \beta ; r \rangle = \frac{1}{p} \sum_{k = 0}^{p-1} 
e^{i \pi \{ (a-b)k^2 + [p(b-a) + 2(\beta - \alpha)]k \} / p}
     \label{inner product p prime}
     \end{eqnarray}
The right-hand side of (\ref{inner product p prime}) can be expressed 
in terms of a generalized quadratic Gauss sum. This leads to 
     \begin{eqnarray}
\langle a \alpha ; r | b \beta ; r \rangle = \frac{1}{p} S(u, v, w)   
     \label{G1}
     \end{eqnarray}
where the Gauss sum $S(u, v, w)$ is given by (\ref{Gauss sum}) with the parameters
     \begin{eqnarray}
     u    =    a - b, \quad v    =    -(a - b)p - 2(\alpha - \beta), \quad w    =    p
     \label{G2}
     \end{eqnarray}
which ensure that $uw+v$ is even. The generalized Gauss sum $S(u, v, w)$ in (\ref{G1})-(\ref{G2}) 
can be calculated from the methods described in \citep{Berndtetal1998}. We thus obtain 
     \begin{eqnarray}
 | \langle a \alpha ; r | b \beta ; r \rangle | = \frac{1}{\sqrt{p}} 
     \label{module du ps BraBrb}
     \end{eqnarray}
for all $a$, $b$, $\alpha$, and $\beta$ in $\mathbb{Z}/p\mathbb{Z}$ with $b \not= a$. This completes the proof. 

Theorem 3 renders feasible to derive in one step the $(p+1)p$ qupits (i.e., qudits with $d=p$ a prime 
integer) of a complete set of $p+1$ MUBs in $\mathbb{C}^p$. The single formula (\ref{aalphar en n bis}) or 
(\ref{eigenvectors of Vra}), giving the $p^2$ vectors 
$| a \alpha ; r \rangle$ or $\phi(a \alpha ; r)$, with $a, \alpha = 0, 1, \ldots, p-1$, of the bases  
$B_{r0}, B_{r1}, \ldots, B_{rp-1}$, is easily codable on a classical computer. 

{\bf Example 6}: The $p=2$ case. For $r=0$, the $p+1 = 3$ MUBs are
     \begin{eqnarray}
B_{00} &:&   \frac{|0\rangle + |1\rangle}{\sqrt{2}}, \quad - \frac{|0\rangle - |1\rangle}{\sqrt{2}} 
		 \nonumber \\
B_{01} &:& i \frac{|0\rangle -i|1\rangle}{\sqrt{2}}, \quad -i\frac{|0\rangle +i|1\rangle}{\sqrt{2}} 
		 \label{MUBs en 2 dim} \\
B_{2}  &:& |0\rangle, \quad |1\rangle 
     \nonumber
     \end{eqnarray}
cf.~(\ref{dim2-1 en jm}), (\ref{dim2-2 en jm}), (\ref{dim2-1}) and (\ref{dim2-2}). The 
global factors $-1$ in $B_{00}$ and $\pm i$ in $B_{01}$ arise from the general formula 
(\ref{aalphar en n bis}); they are irrelevant for quantum information and can be omitted.

{\bf Example 7}: The $p=3$ case. For $r=0$, the $p+1 = 4$ MUBs are
     \begin{eqnarray}
B_{00} &:&   \frac{   |0\rangle +    |1\rangle + |2\rangle}{\sqrt{3}}, \quad 
             \frac{q^2|0\rangle + q  |1\rangle + |2\rangle}{\sqrt{3}}, \quad 
             \frac{q  |0\rangle + q^2|1\rangle + |2\rangle}{\sqrt{3}} 
		 \nonumber \\
B_{01} &:&   \frac{q  |0\rangle + q  |1\rangle + |2\rangle}{\sqrt{3}}, \quad 
             \frac{   |0\rangle + q^2|1\rangle + |2\rangle}{\sqrt{3}}, \quad 
             \frac{q^2|0\rangle +    |1\rangle + |2\rangle}{\sqrt{3}} 
		 \nonumber \\
		 \label{MUBs en 3 dim} \\
B_{02} &:&   \frac{q^2|0\rangle + q^2|1\rangle + |2\rangle}{\sqrt{3}}, \quad 
             \frac{q  |0\rangle +    |1\rangle + |2\rangle}{\sqrt{3}}, \quad 
             \frac{   |0\rangle + q  |1\rangle + |2\rangle}{\sqrt{3}} 
		 \nonumber \\		 
B_{3}  &:& |0\rangle, \quad |1\rangle, \quad |2\rangle 		 
     \nonumber
     \end{eqnarray}
with $q = e^{i 2 \pi / 3}$, cf.~(\ref{62}), (\ref{63}), (\ref{64}), (\ref{010}), 
(\ref{110}) and (\ref{210}). 

As a simple consequence of Theorem 3, we get the following corollary which can be 
derived by combining Theorem 3 with Eq.~(\ref{matrix Fra and inner product}). 

{\bf Corollary 2}. {\it For $d=p$, with $p$ a prime number, the $p \times p$ matrix 
$({\bf F_{ra}})^{\dagger} {\bf F_{rb}}$ with $b \not= a$ ($a, b = 0, 1, \ldots, p-1$) 
is a generalized Hadamard matrix. 
}

Going back to arbitrary $d$, it is to be noted that for a fixed value of $r$, the $d+1$ bases 
$B_{r0}$, $B_{r1}$, $\ldots$, $B_{rd-1}$, $B_{d}$ do not provide in general a complete set of 
$d+1$ MUBs even in the case where $d$ is a power $p^e$ with $e \geq 2$ of a prime integer $p$. 
However, it is possible to show \citep{Kibler2009} that the bases $B_{0a}$, $B_{0 a \oplus 1}$ 
and $B_d$ are three MUBs in ${\mathbb{C}}^d$, 
in agreement with $N_{\scriptscriptstyle MUB} \geq 3$. Therefore for $d$ arbitrary, 
given two Hadamard matrices ${\bf F_{ra}}$ and ${\bf F_{sb}}$, the product 
${\bf F_{ra}}^{\dagger} {\bf F_{sb}}$ is not in general a Hadamard matrix. 

In the case where $d$ is a power $p^e$ with $e \geq 2$ of a prime integer $p$, tensor products of 
the unbiased bases $B_{r0}$, $B_{r1}$, $\ldots$, $B_{rp-1}$ can be used for generating $p^e+1$
MUBs in dimension $d = p^e$. This can be illustrated with the following example. 

{\bf Example 8}: The $d=2^2$ case. This case corresponds to a spin $j = 3/2$. The application of 
(\ref{j alpha r a in terms of jm}) or (\ref{aalphar en n bis}) yields four bases 
$B_{0a}$ ($a = 0, 1, 2, 3$). As a point of fact, the five bases $B_{00}$, $B_{01}$, $B_{02}$, $B_{03}$ and $B_4$ 
do not form a complete set of $d+1 = 5$ MUBs ($d=4$ is not a prime number). Nevertheless, it is possible to find 
five MUBs because $d = 2^2$ is the power of a prime number. This can be 
achieved by replacing the space $\epsilon(4)$ spanned by 
   \begin{eqnarray}
B_4 = \{ | 3/2 , m \rangle : m = 3/2, 1/2, -1/2, -3/2 \} \quad {\rm or} \quad \{ | n \rangle : n = 0, 1, 2, 3 \}
 	 \end{eqnarray}
by the tensor product space $\epsilon(2) \otimes \epsilon(2)$ spanned by the canonical or computational basis 
   \begin{eqnarray}
B_2 \otimes B_2 = \{ | 0 \rangle \otimes | 0 \rangle, | 0 \rangle \otimes | 1 \rangle, | 1 \rangle \otimes | 0 \rangle, | 1 \rangle \otimes | 1 \rangle \}
 	 \end{eqnarray}
The space $\epsilon(2) \otimes \epsilon(2)$ 
is associated with the coupling of two spin angular momenta 
$j_1 = 1/2$ and $j_2 = 1/2$ or two qubits (in the vector $u \otimes v$, $u$ and $v$ correspond to 
$j_1$       and $j_2$, respectively). Four of the five MUBs for $d=4$ can be constructed from the direct products 
   \begin{eqnarray}
|a b : \alpha \beta \rangle    =    |a \alpha ; 0 \rangle \otimes |b \beta ; 0 \rangle
 	 \end{eqnarray}
which are eigenvectors of the operators 
   \begin{eqnarray}
w_{ab}    =    v_{0a} \otimes v_{0b}
 	 \end{eqnarray} 
 (the operators $v_{0a}$ and $v_{0b}$ refer to the two spaces $\epsilon(2)$,  
 the vectors of type $|a \alpha ; 0 \rangle$ and $|b \beta ; 0 \rangle$ are 
 given by the master formula (\ref{aalphar en n bis}) for $d=2$). Obviously, the set  	 
   	\begin{eqnarray}
B_{0a0b} = \{ |a b : \alpha \beta \rangle : \alpha, \beta = 0, 1 \}
   	\end{eqnarray} 
is an orthonormal basis in ${\mathbb{C}}^4$. It is evident that $B_{0000}$ and $B_{0101}$ are two unbiased bases 
since the modulus of the inner product of 
$|0 0 : \alpha \beta \rangle$ by $|1 1 : \alpha' \beta' \rangle$ is 
   	\begin{eqnarray}
| \langle 0 0 : \alpha \beta | 1 1 : \alpha' \beta' \rangle | =   	
| \langle 0 \alpha ; 0 | 1 \alpha' ; 0 \rangle 
  \langle 0 \beta  ; 0 | 1 \beta'  ; 0 \rangle | = \frac{1}{\sqrt{2}} \frac{1}{\sqrt{2}}= \frac{1}{\sqrt{4}}
   	\end{eqnarray}    	
A similar result holds for the two bases $B_{0001}$ and $B_{0100}$. However, 
the four bases $B_{0000}$, $B_{0101}$, $B_{0001}$ and $B_{0100}$ are not 
mutually unbiased. A possible way to overcome this no-go result is to keep 
the bases $B_{0000}$ and $B_{0101}$ intact and to re-organize the vectors 
inside the bases $B_{0001}$ and $B_{0100}$ in order to obtain four MUBs. We 
are thus left with four bases 
   	\begin{eqnarray}
W_{00} \equiv B_{0000}, \quad W_{11} \equiv B_{0101}, \quad W_{01}, \quad W_{10} 
   	\end{eqnarray}   
which together with the computational basis $B_4$ give five MUBs. Specifically, we have 
   	\begin{eqnarray}
W_{00} & = & \{         |0 0 : \alpha \beta \rangle : \alpha, \beta = 0, 1 \}  \\
W_{11} & = & \{         |1 1 : \alpha \beta \rangle : \alpha, \beta = 0, 1 \}  \\
W_{01} & = & \{ \lambda |0 1 : \alpha \beta \rangle + 
                    \mu |0 1 : \alpha \oplus 1 \beta \oplus 1 \rangle : \alpha, \beta = 0, 1 \}  \\
W_{10} & = & \{ \lambda |1 0 : \alpha \beta \rangle + 
                    \mu |1 0 : \alpha \oplus 1 \beta \oplus 1 \rangle : \alpha, \beta = 0, 1 \}  
    \label{les quatre W}
   	\end{eqnarray} 
where
   	\begin{eqnarray}
\lambda = \frac{1-i}{2}, \quad \mu = \frac{1+i}{2}
   	\end{eqnarray} 
As a r\'esum\'e, only 
two formulas are necessary for obtaining the $d^2 = 16$ vectors for the bases $W_{ab}$, namely 
   	  \begin{eqnarray}
W_{00}, W_{11} & : & | a a : \alpha \beta \rangle 
\label{W00W11} \\
W_{01}, W_{10} & : & \lambda |a a \oplus 1 : \alpha       \beta \rangle + 
                                                                 \mu |a a \oplus 1 : \alpha \oplus 1 \beta \oplus 1 \rangle 
      \label{W01W10}
   	  \end{eqnarray} 
for all $a, \alpha$ and $\beta$ in ${\mathbb{Z}}/2{\mathbb{Z}}$. The five MUBs are listed 
below as state vectors and column vectors with
   	\begin{eqnarray}
\vert 0 \rangle \equiv \vert j=1/2 , m = 1/2 \rangle \ {\rm or} \  
			\begin{pmatrix}
1      \cr
0      \cr
			\end{pmatrix}, \quad 
\vert 1 \rangle \equiv \vert j=1/2 , m =-1/2 \rangle \ {\rm or} \  
\begin{pmatrix}
0      \cr
1      \cr
		\end{pmatrix}       
		\label{vvvvv}
   	\end{eqnarray} 

{The canonical basis}: 
   \begin{eqnarray}
| 0 \rangle \otimes | 0 \rangle, \quad
| 0 \rangle \otimes | 1 \rangle, \quad
| 1 \rangle \otimes | 0 \rangle, \quad
| 1 \rangle \otimes | 1 \rangle   
   \nonumber 
 	 \end{eqnarray} 
or in column vectors
	\begin{eqnarray}
	\begin{pmatrix}
1 \cr
0 \cr
0 \cr
0 \cr
	\end{pmatrix}, \quad
	\begin{pmatrix}
0 \cr
1 \cr
0 \cr
0 \cr
	\end{pmatrix}, \quad
	\begin{pmatrix}
0 \cr
0 \cr
1 \cr
0 \cr
	\end{pmatrix}, \quad
	\begin{pmatrix}
0 \cr
0 \cr
0 \cr
1 \cr
	\end{pmatrix}       
	\label{canonical basis CCCC08}
	\end{eqnarray}

{The $W_{00}$ basis}:
\begin{eqnarray}
&& | 0 0 : 0 0 \rangle =+\frac{| 0 \rangle + | 1 \rangle}{\sqrt{2}}  \otimes 
                         \frac{| 0 \rangle + | 1 \rangle}{\sqrt{2}}, \quad
   | 0 0 : 0 1 \rangle =-\frac{| 0 \rangle + | 1 \rangle}{\sqrt{2}}  \otimes 
                         \frac{| 0 \rangle - | 1 \rangle}{\sqrt{2}}  \nonumber \\
&& | 0 0 : 1 0 \rangle =-\frac{| 0 \rangle - | 1 \rangle}{\sqrt{2}}  \otimes 
                         \frac{| 0 \rangle + | 1 \rangle}{\sqrt{2}}, \quad
   | 0 0 : 1 1 \rangle =+\frac{| 0 \rangle - | 1 \rangle}{\sqrt{2}}  \otimes 
                         \frac{| 0 \rangle - | 1 \rangle}{\sqrt{2}}  \nonumber
\end{eqnarray}
or in developed form
\begin{eqnarray}
&& | 0 0 : 0 0 \rangle  =  + 
\frac{1}{2}(| 0 \rangle \otimes | 0 \rangle + | 0 \rangle \otimes | 1 \rangle + | 1 \rangle \otimes | 0 \rangle + | 1 \rangle \otimes | 1 \rangle) \nonumber \\ 
&& | 0 0 : 0 1 \rangle  =  - 
\frac{1}{2}(| 0 \rangle \otimes | 0 \rangle - | 0 \rangle \otimes | 1 \rangle + | 1 \rangle \otimes | 0 \rangle - | 1 \rangle \otimes | 1 \rangle) \nonumber \\ 
&& | 0 0 : 1 0 \rangle  =  -  
\frac{1}{2}(| 0 \rangle \otimes | 0 \rangle + | 0 \rangle \otimes | 1 \rangle - | 1 \rangle \otimes | 0 \rangle - | 1 \rangle \otimes | 1 \rangle) \nonumber \\ 
&& | 0 0 : 1 1 \rangle  =  + 
\frac{1}{2}(| 0 \rangle \otimes | 0 \rangle - | 0 \rangle \otimes | 1 \rangle - | 1 \rangle \otimes | 0 \rangle + | 1 \rangle \otimes | 1 \rangle) \nonumber        
\end{eqnarray}
or in column vectors
	\begin{eqnarray}
\frac{1}{2} 
		\begin{pmatrix}
1 \cr
1 \cr
1 \cr
1 \cr
		\end{pmatrix}, \quad
-\frac{1}{2} 
		\begin{pmatrix}
1  \cr
-1 \cr
1  \cr
-1 \cr
		\end{pmatrix}, \quad
-\frac{1}{2} 
		\begin{pmatrix}
1  \cr
1  \cr
-1 \cr
-1 \cr
		\end{pmatrix}, \quad
\frac{1}{2} 
		\begin{pmatrix}
1  \cr
-1 \cr
-1 \cr
1  \cr
	\end{pmatrix}       
	\label{W00 basis CCCC08}
	\end{eqnarray}

{The $W_{11}$ basis}:
\begin{eqnarray}
&& | 1 1 : 0 0 \rangle =-\frac{| 0 \rangle -i| 1 \rangle}{\sqrt{2}}  \otimes 
                         \frac{| 0 \rangle -i| 1 \rangle}{\sqrt{2}}, \quad
   | 1 1 : 0 1 \rangle =+\frac{| 0 \rangle -i| 1 \rangle}{\sqrt{2}}  \otimes 
                         \frac{| 0 \rangle +i| 1 \rangle}{\sqrt{2}}  \nonumber \\
&& | 1 1 : 1 0 \rangle =+\frac{| 0 \rangle +i| 1 \rangle}{\sqrt{2}}  \otimes 
                         \frac{| 0 \rangle -i| 1 \rangle}{\sqrt{2}}, \quad
   | 1 1 : 1 1 \rangle =-\frac{| 0 \rangle +i| 1 \rangle}{\sqrt{2}}  \otimes 
                         \frac{| 0 \rangle +i| 1 \rangle}{\sqrt{2}}  \nonumber
\end{eqnarray}
or in developed form
\begin{eqnarray}
&& | 1 1 : 0 0 \rangle  =  -
\frac{1}{2}(| 0 \rangle \otimes | 0 \rangle - i | 0 \rangle \otimes | 1 \rangle - i | 1 \rangle \otimes | 0 \rangle - | 1 \rangle \otimes | 1 \rangle) \nonumber \\
&& | 1 1 : 0 1 \rangle  =  +
\frac{1}{2}(| 0 \rangle \otimes | 0 \rangle + i | 0 \rangle \otimes | 1 \rangle - i | 1 \rangle \otimes | 0 \rangle + | 1 \rangle \otimes | 1 \rangle) \nonumber \\ 
&& | 1 1 : 1 0 \rangle  =  +
\frac{1}{2}(| 0 \rangle \otimes | 0 \rangle - i | 0 \rangle \otimes | 1 \rangle + i | 1 \rangle \otimes | 0 \rangle + | 1 \rangle \otimes | 1 \rangle) \nonumber \\ 
&& | 1 1 : 1 1 \rangle  =  -
\frac{1}{2}(| 0 \rangle \otimes | 0 \rangle + i | 0 \rangle \otimes | 1 \rangle + i | 1 \rangle \otimes | 0 \rangle - | 1 \rangle \otimes | 1 \rangle) \nonumber 
\end{eqnarray}
or in column vectors
	\begin{eqnarray}
-\frac{1}{2} 
	\begin{pmatrix}
1  \cr
-i \cr
-i \cr
-1 \cr 
  \end{pmatrix}, \quad  	
\frac{1}{2} 
	\begin{pmatrix}
1  \cr
i  \cr
-i \cr
1  \cr
  \end{pmatrix}, \quad
\frac{1}{2} 
  \begin{pmatrix}
1  \cr
-i \cr
i  \cr
1  \cr
	\end{pmatrix}, \quad    
- \frac{1}{2} 
	\begin{pmatrix}
1  \cr
i  \cr
i  \cr
-1  \cr
	\end{pmatrix}   
	\label{W11 basis CCCC08}
	\end{eqnarray}

{The $W_{01}$ basis}:
\begin{eqnarray}
&& \lambda | 0 1 : 0 0 \rangle + \mu | 0 1 : 1 1 \rangle  =  
+\mu      \frac{| 0 \rangle + | 1 \rangle}{\sqrt{2}} \otimes 
          \frac{| 0 \rangle -i| 1 \rangle}{\sqrt{2}}  
-\lambda  \frac{| 0 \rangle - | 1 \rangle}{\sqrt{2}} \otimes    
          \frac{| 0 \rangle +i| 1 \rangle}{\sqrt{2}} \nonumber \\
&& \mu | 0 1 : 0 0 \rangle + \lambda | 0 1 : 1 1 \rangle  =  
-\lambda  \frac{| 0 \rangle + | 1 \rangle}{\sqrt{2}} \otimes 
          \frac{| 0 \rangle -i| 1 \rangle}{\sqrt{2}}  
+\mu      \frac{| 0 \rangle - | 1 \rangle}{\sqrt{2}} \otimes    
          \frac{| 0 \rangle +i| 1 \rangle}{\sqrt{2}} \nonumber \\
&& \lambda | 0 1 : 0 1 \rangle + \mu | 0 1 : 1 0 \rangle  = 
-\mu      \frac{| 0 \rangle + | 1 \rangle}{\sqrt{2}} \otimes 
          \frac{| 0 \rangle +i| 1 \rangle}{\sqrt{2}}  
+\lambda  \frac{| 0 \rangle - | 1 \rangle}{\sqrt{2}} \otimes    
          \frac{| 0 \rangle -i| 1 \rangle}{\sqrt{2}} \nonumber \\                                                 
&& \mu | 0 1 : 0 1 \rangle + \lambda | 0 1 : 1 0 \rangle  = 
+\lambda  \frac{| 0 \rangle + | 1 \rangle}{\sqrt{2}} \otimes 
          \frac{| 0 \rangle +i| 1 \rangle}{\sqrt{2}}  
-\mu      \frac{| 0 \rangle - | 1 \rangle}{\sqrt{2}} \otimes    
          \frac{| 0 \rangle -i| 1 \rangle}{\sqrt{2}} \nonumber 
\end{eqnarray}          
or in developed form
\begin{eqnarray}
&& \lambda | 0 1 : 0 0 \rangle + \mu | 0 1 : 1 1 \rangle  =  + 
\frac{i}{2}(| 0 \rangle \otimes | 0 \rangle - | 0 \rangle \otimes | 1 \rangle - i | 1 \rangle \otimes | 0 \rangle - i | 1 \rangle \otimes | 1 \rangle) \nonumber \\ 
&& \mu | 0 1 : 0 0 \rangle + \lambda | 0 1 : 1 1 \rangle  =  + 
\frac{i}{2}(| 0 \rangle \otimes | 0 \rangle + | 0 \rangle \otimes | 1 \rangle + i | 1 \rangle \otimes | 0 \rangle - i | 1 \rangle \otimes | 1 \rangle) \nonumber \\ 
&& \lambda | 0 1 : 0 1 \rangle + \mu | 0 1 : 1 0 \rangle  =  - 
\frac{i}{2}(| 0 \rangle \otimes | 0 \rangle + | 0 \rangle \otimes | 1 \rangle - i | 1 \rangle \otimes | 0 \rangle + i | 1 \rangle \otimes | 1 \rangle) \nonumber \\ 
&& \mu | 0 1 : 0 1 \rangle + \lambda | 0 1 : 1 0 \rangle  =  - 
\frac{i}{2}(| 0 \rangle \otimes | 0 \rangle - | 0 \rangle \otimes | 1 \rangle + i | 1 \rangle \otimes | 0 \rangle + i | 1 \rangle \otimes | 1 \rangle) \nonumber 
\end{eqnarray}
or in column vectors
	\begin{eqnarray}
  \frac{i}{2} \begin{pmatrix}
1  \cr
-1 \cr
-i \cr
-i \cr
  \end{pmatrix}, \quad
  \frac{i}{2} \begin{pmatrix}
1  \cr
1  \cr
i  \cr
-i \cr
	\end{pmatrix}, \quad       
  -\frac{i}{2} \begin{pmatrix}
1  \cr
1  \cr
-i \cr
i  \cr
\end{pmatrix}, \quad
  -\frac{i}{2} \begin{pmatrix}
1  \cr
-1 \cr
i  \cr
i  \cr
\end{pmatrix}, 
	\label{W01 basis CCCC08}
	\end{eqnarray}

{The $W_{10}$ basis}:
\begin{eqnarray}
&& \lambda | 1 0 : 0 0 \rangle + \mu | 1 0 : 1 1 \rangle  =  
+\mu      \frac{| 0 \rangle -i| 1 \rangle}{\sqrt{2}} \otimes 
          \frac{| 0 \rangle + | 1 \rangle}{\sqrt{2}}  
-\lambda  \frac{| 0 \rangle +i| 1 \rangle}{\sqrt{2}} \otimes    
          \frac{| 0 \rangle - | 1 \rangle}{\sqrt{2}} \nonumber \\
&& \mu | 1 0 : 0 0 \rangle + \lambda | 1 0 : 1 1 \rangle  =  
-\lambda  \frac{| 0 \rangle -i| 1 \rangle}{\sqrt{2}} \otimes 
          \frac{| 0 \rangle + | 1 \rangle}{\sqrt{2}}  
+\mu      \frac{| 0 \rangle +i| 1 \rangle}{\sqrt{2}} \otimes    
          \frac{| 0 \rangle - | 1 \rangle}{\sqrt{2}} \nonumber \\
&& \lambda | 1 0 : 0 1 \rangle + \mu | 1 0 : 1 0 \rangle  = 
-\mu      \frac{| 0 \rangle -i| 1 \rangle}{\sqrt{2}} \otimes 
          \frac{| 0 \rangle - | 1 \rangle}{\sqrt{2}}  
+\lambda  \frac{| 0 \rangle +i| 1 \rangle}{\sqrt{2}} \otimes    
          \frac{| 0 \rangle + | 1 \rangle}{\sqrt{2}} \nonumber \\ 
&& \mu | 1 0 : 0 1 \rangle + \lambda | 1 0 : 1 0 \rangle  = 
+\lambda  \frac{| 0 \rangle -i| 1 \rangle}{\sqrt{2}} \otimes 
          \frac{| 0 \rangle - | 1 \rangle}{\sqrt{2}}  
-\mu      \frac{| 0 \rangle +i| 1 \rangle}{\sqrt{2}} \otimes    
          \frac{| 0 \rangle + | 1 \rangle}{\sqrt{2}} \nonumber 
\end{eqnarray}          
or in developed form
\begin{eqnarray}
&& \lambda | 1 0 : 0 0 \rangle + \mu | 1 0 : 1 1 \rangle  = +
\frac{i}{2}(| 0 \rangle \otimes | 0 \rangle - i | 0 \rangle \otimes | 1 \rangle - | 1 \rangle \otimes | 0 \rangle - i | 1 \rangle \otimes | 1 \rangle) \nonumber \\
&& \mu | 1 0 : 0 0 \rangle + \lambda | 1 0 : 1 1 \rangle  = + 
\frac{i}{2}(| 0 \rangle \otimes | 0 \rangle + i | 0 \rangle \otimes | 1 \rangle + | 1 \rangle \otimes | 0 \rangle - i | 1 \rangle \otimes | 1 \rangle) \nonumber \\ 
&& \lambda | 1 0 : 0 1 \rangle + \mu | 1 0 : 1 0 \rangle  = -
\frac{i}{2}(| 0 \rangle \otimes | 0 \rangle + i | 0 \rangle \otimes | 1 \rangle - | 1 \rangle \otimes | 0 \rangle + i | 1 \rangle \otimes | 1 \rangle) \nonumber \\ 
&& \mu | 1 0 : 0 1 \rangle + \lambda | 1 0 : 1 0 \rangle  = - 
\frac{i}{2}(| 0 \rangle \otimes | 0 \rangle - i | 0 \rangle \otimes | 1 \rangle + | 1 \rangle \otimes | 0 \rangle + i | 1 \rangle \otimes | 1 \rangle) \nonumber  
\end{eqnarray}
or in column vectors
	\begin{eqnarray}
\frac{i}{2} 
	\begin{pmatrix}
1  \cr
-i \cr
-1 \cr
-i \cr
	\end{pmatrix}, \quad
\frac{i}{2} 
	\begin{pmatrix}
1  \cr
i  \cr
1  \cr
-i \cr
	\end{pmatrix}, \quad
-\frac{i}{2} 
	\begin{pmatrix}
1  \cr
i  \cr
-1 \cr
i  \cr
	\end{pmatrix}, \quad
-\frac{i}{2} 
	\begin{pmatrix}
1  \cr
-i \cr
1  \cr
i  \cr
	\end{pmatrix}	
	\label{W10 basis CCCC08}
	\end{eqnarray}

The five preceding bases are of central importance in quantum information 
for expressing any ququart or quartic (corresponding to $d=4$) in terms 
of qudits (corresponding to $d=2$). It is to be noted that the vectors 
of the $W_{00}$ and $W_{11}$ bases are not intricated (i.e., each vector 
is the direct product of two vectors) while the vectors of the $W_{01}$ 
and $W_{10}$ bases are intricated (i.e., each vector is not the direct 
product of two vectors). To be more precise, the degree of intrication 
of the state vectors for the bases $W_{00}$, $W_{11}$, $W_{01}$ and $W_{10}$ 
can be determined in the following way. In arbitrary dimension $d$, let 
\begin{eqnarray}
| \Phi \rangle = \sum_{k = 0}^{d-1} \sum_{l = 0}^{d-1} a_{kl} | k \rangle \otimes | l \rangle
\end{eqnarray}
be a double qudit state vector. Then, it can be shown that the 
determinant of the $d \times d$ matrix $A = (a_{kl})$ satisfies 
\begin{eqnarray}
0 \leq |\det A| \leq \frac{1}{\sqrt{d^d}}
\end{eqnarray}
as discussed in \citep{Albouy2009}. The case $\det A = 0$ 
corresponds to the absence of \emph{global} intrication while the case 
\begin{eqnarray}
|\det A| = \frac{1}{\sqrt{d^d}}
\end{eqnarray} 
corresponds to a maximal intrication. As an illustration, we obtain that all the 
state vectors for $W_{00}$ and $W_{11}$ are not intricated and that all the state 
vectors for $W_{01}$ and $W_{10}$ are maximally intricated.

Generalization of (\ref{W00W11}) and (\ref{W01W10}) can be obtained in more complicated situations 
(two qupits, three qubits, \ldots). The generalization of (\ref{W00W11}) is immediate. The generalization 
of (\ref{W01W10}) can be achieved by taking linear combinations of vectors such that each linear combination 
is made of vectors corresponding to the same eigenvalue of the relevant tensor product of operators of type 
$v_{0a}$.

 \subsection{Mutually unbiased bases and Lie agebras}

\subsubsection{Generalized Pauli matrices}

We now examine the interest for quantum information of the Weyl pair $({\bf X} , {\bf Z})$ introduced 
in section 3.1.4. The linear operators corresponding to the matrices ${\bf X}$ and ${\bf Z}$ are known 
in quantum information and quantum computing as shift and clock operators, respectively. (Note however 
that for each of the operators $x$ and $z$, the {\it shift} or {\it clock} character depends on which 
state the operator acts. The qualification adopted in quantum information and quantum computing corresponds 
to the action of $x$ and $z$ on the computational basis $B_d$.) For $d$ arbitrary, they are at the root of 
the Pauli group $P_d$, a finite subgroup of $U(d)$ (see section 3.1.4). The normaliser of $P_d$ in $U(d)$ 
is a Clifford-type group in $d$ dimensions noted $Cl_d$. More precisely, $Cl_d$ is the set 
$\{ {\bf U} \in U(d) | {\bf U} P_d {\bf U}^{\dagger} = P_d \}$ endowed with matrix multiplication 
(the elements of $P_d$ being expressed in terms of the matrices ${\bf X}$ and ${\bf Z}$). The Pauli 
group $P_d$, as well as any other invariant subgroup of $Cl_d$, is of considerable importance for 
describing quantum errors and quantum fault tolerance in quantum computing 
(see \citep{HavlicekSaniga2008,Planat2010,PlanatKibler2010} 
and references therein for recent geometrical approaches to the Pauli group). These concepts are very 
important in the case of $n$-qubit systems (corresponding to $d = 2^n$). 

The Weyl pair (${\bf X} , {\bf Z}$) turns out to be an integrity basis for generating 
the set $\{ {\bf X}^a {\bf Z}^b : a,b \in \mathbb{Z}/d\mathbb{Z}\}$ of $d^2$ generalized 
Pauli matrices in $d$ dimensions (see for instance 
\citep{Gottesmanetal2001,Lawrenceetal2002,Bandyopadhyayetal2002,PittengerRubin2004,Kibler2008} 
in the context of MUBs and \citep{Tolar1984,BalianItzykson1986,PateraZassenhaus1988} 
in group-theoretical contexts). As seen in section 3.1.4, the latter set constitutes a basis for 
the Lie algebra of the linear group $GL(d , \mathbb{C})$ (or its unitary restriction $U(d)$) with 
respect to the commutator law. Let us give two examples of these important generalized Pauli matrices.

{\bf Example 9}: The $d = 2$ case. For $d=2 \Leftrightarrow j=1/2$ ($\Rightarrow q = -1$), the matrices 
of the four operators $u_{ab}$ with $a, b = 0,1$ are 
		\begin{eqnarray}
{\bf I}_2 = {\bf X}^0 {\bf Z}^0, \quad 
{\bf X}   = {\bf X}^1 {\bf Z}^0, \quad 
{\bf Y}   = {\bf X}^1 {\bf Z}^1, \quad 
{\bf Z}   = {\bf X}^0 {\bf Z}^1   
		\end{eqnarray}
or in terms of the matrices ${\bf V_{0a}}$
		\begin{eqnarray}
{\bf I}_2          = ({\bf V_{00}})^2, \quad 
{\bf X}            =  {\bf V_{00}},    \quad   
{\bf Y}            =  {\bf V_{01}},    \quad 
{\bf Z}            = ({\bf V_{00}})^{\dagger} {\bf V_{01}} 
		\end{eqnarray}
In detail, we get 
		\begin{eqnarray}
{\bf I}_2                 
		= \begin{pmatrix}
  1     &0   \cr
  0     &1   \cr
		\end{pmatrix}, \quad 
{\bf X}                      
		= \begin{pmatrix}
  0     &1   \cr
  1     &0   \cr
		\end{pmatrix}, \quad  
{\bf Y}                                
		= \begin{pmatrix}
  0     &-1  \cr
  1     &0   \cr
		\end{pmatrix}, \quad 
{\bf Z}   
		= \begin{pmatrix}
  1     &0    \cr
  0     &-1   \cr
		\end{pmatrix} 
		\end{eqnarray}		
Alternatively, we have
		\begin{eqnarray}
{\bf I}_2 =     \sigma_0, \quad		
{\bf X}   =     \sigma_x, \quad   
{\bf Y}   = - i \sigma_y, \quad 
{\bf Z}   =     \sigma_z
		\end{eqnarray}
in terms of the usual (Hermitian and unitary) Pauli matrices 
$\sigma_0$, $\sigma_x$, $\sigma_y$ and $\sigma_z$
\begin{eqnarray}
\sigma_0    =    
		\begin{pmatrix}
  1     &0    \cr
  0     &1    \cr
		\end{pmatrix}, \quad 
\sigma_x    =    
		\begin{pmatrix}
  0     &1   \cr
  1     &0   \cr
		\end{pmatrix}, \quad 
\sigma_y    =    
		\begin{pmatrix}
  0     &-i  \cr
  i     &0   \cr
		\end{pmatrix}, \quad 
\sigma_z    =    
		\begin{pmatrix}
  1     &0    \cr
  0     &-1   \cr
		\end{pmatrix}
		\end{eqnarray}
The approach 
developed here leads to generalized Pauli matrices in dimension 2 
that differ from the usual Pauli matrices. This is the price one has to pay in order 
to get a systematic generalization of Pauli matrices in arbitrary dimension. 
It should be observed that the commutation and anti-commutation 
relations given by (\ref{com new}) and (\ref{anticom new}) with $d=2$ correspond to the well-known 
commutation and anti-commutation relations for the usual Pauli matrices transcribed 
in the normalization 
${\bf X}^1 {\bf Z}^0 =    \sigma_x$, 
${\bf X}^1 {\bf Z}^1 = -i \sigma_y$,  
${\bf X}^0 {\bf Z}^1 =    \sigma_z$.

From a group-theoretical point of view, the matrices ${\bf I}_2$, {\bf X}, {\bf Y} and {\bf Z} can be 
considered as generators of the group $U(2)$. On the other hand, the Pauli group $P_2$ contains eight 
elements; due to the factor $-i$ in ${\bf Y} = -i \sigma_y$, the group $P_2$ is isomorphic to the group 
of hyperbolic quaternions rather than to the group of ordinary quaternions.

In terms of column vectors, the vectors of the bases $B_{00}$, $B_{01}$ and $B_{2}$ (see (\ref{MUBs en 2 dim})) 
are eigenvevtors of $\sigma_x$, $\sigma_y$ and $\sigma_z$, respectively (for each matrix the eigenvalues are 
$1$ and $-1$).

{\bf Example 10}: The $d=3$ case. For $d=3 \Leftrightarrow j=1$ ($\Rightarrow q = e^{{i} 2 \pi / 3}$), the 
matrices of the nine operators $u_{ab}$ with $a, b = 0,1,2$, viz., 
     \begin{eqnarray}
&& 	{\bf X}^0 {\bf Z}^0 = {\bf I_3},         \quad 
		{\bf X}^1 {\bf Z}^0 = {\bf X},           \quad
		{\bf X}^2 {\bf Z}^0 = {\bf X}^2          \nonumber \\	
&&	{\bf X}^0 {\bf Z}^1 = {\bf Z},           \quad
		{\bf X}^0 {\bf Z}^2 = {\bf Z}^2,         \quad
   	{\bf X}^1 {\bf Z}^1 = {\bf X} {\bf Z}              \\ 
&&	{\bf X}^2 {\bf Z}^2,                     \quad
		{\bf X}^2 {\bf Z}^1 = {\bf X}^2 {\bf Z}, \quad  
		{\bf X}^1 {\bf Z}^2 = {\bf X} {\bf Z}^2  \nonumber
     \end{eqnarray}
are 
     \begin{eqnarray}
{\bf I_3} = 
		\begin{pmatrix}
  1     &0     &0   \cr
  0     &1     &0   \cr
  0     &0     &1   \cr
		\end{pmatrix}, \quad 
{\bf X} = 
		\begin{pmatrix}
  0     &1     &0   \cr
  0     &0     &1   \cr
  1     &0     &0   \cr
		\end{pmatrix}, \quad 
{\bf X}^2 = 
		\begin{pmatrix}
  0     &0     &1   \cr
  1     &0     &0   \cr
  0     &1     &0   \cr
		\end{pmatrix} 
\nonumber
     \end{eqnarray}
     \begin{eqnarray}
{\bf Z} = 
		\begin{pmatrix}
  1     &0     &0     \cr
  0     &q     &0     \cr
  0     &0     &q^2   \cr
		\end{pmatrix}, \quad 
{\bf Z}^2 = 
		\begin{pmatrix}
  1     &0       &0   \cr
  0     &q^2     &0   \cr
  0     &0       &q   \cr
		\end{pmatrix}, \quad 
{\bf X} {\bf Z} = 
		\begin{pmatrix}
  0     &q     &0     \cr
  0     &0     &q^2   \cr
  1     &0     &0     \cr
		\end{pmatrix}
\label{neuf matrices}
     \end{eqnarray}
     \begin{eqnarray}
{\bf X}^2 {\bf Z}^2 = 
		\begin{pmatrix}
  0     &0       &q     \cr
  1     &0       &0     \cr
  0     &q^2     &0     \cr
		\end{pmatrix}, \quad 
{\bf X}^2 {\bf Z} =  
		\begin{pmatrix}
  0     &0     &q^2     \cr
  1     &0     &0       \cr
  0     &q     &0       \cr
		\end{pmatrix}, \quad 
{\bf X} {\bf Z}^2 =  
		\begin{pmatrix}
  0     &q^2     &0     \cr
  0     &0       &q     \cr
  1     &0       &0     \cr
		\end{pmatrix}
\nonumber
     \end{eqnarray}
The generalized Pauli matrices (\ref{neuf matrices}) differ from the Gell-Mann matrices used in 
elementary particle physics. They constitute another extension of the Pauli matrices in dimension 
$d = 3$ of interest for the Lie group $U(3)$ and the Pauli group $P_3$.  

In terms of column vectors, the vectors of the bases $B_{00}$, $B_{01}$, $B_{02}$ and $B_{3}$ 
(see (\ref{MUBs en 3 dim})) are eigenvectors of ${\bf X}$, ${\bf X}{\bf Z}$, ${\bf X}{\bf Z}^2$ 
and ${\bf Z}$, respectively (for each matrix the eigenvalues are $1$, $q$ and $q^2$).

\subsubsection{MUBs and the special linear group}

In the case where $d$ is a prime integer or a power of a prime integer, it is known that the 
set $\{ {\bf X}^a{\bf Z}^b : a, b = 0, 1, \ldots, d-1 \}$ of cardinality $d^2$ can be partitioned into 
$d+1$ subsets containing each $d-1$ commuting matrices (cf.~\citep{Bandyopadhyayetal2002}). Let us give an 
example before going to the case where $d$ is an arbitrary prime number.

{\bf Example 11}: The $d=5$ case. For $d=5$, we have the six following sets of four commuting matrices
           \begin{eqnarray}  	   
&& {\cal V}_0       =   \{  01 ,  02 ,  03 ,  04  \}, \quad	   	   
   {\cal V}_1       =   \{  10 ,  20 ,  30 ,  40  \} 
   \nonumber \\
&& {\cal V}_2       =   \{  11 ,  22 ,  33 ,  44  \}, \quad
   {\cal V}_3       =   \{  12 ,  24 ,  31 ,  43  \} 
             \\
&& {\cal V}_4       =   \{  13 ,  21 ,  34 ,  42  \}, \quad
   {\cal V}_5       =   \{  14 ,  23 ,  32 ,  41  \} 
  \nonumber
          \end{eqnarray} 
where $ab$ is used as an abbreviation of ${\bf X}^a {\bf Z}^b$.

{\bf Proposition 16}. {\it For $d=p$ with $p$ a prime integer, the $p+1$ sets of $p-1$ commuting matrices 
are easily seen to be                 
           \begin{eqnarray}  
{\cal V}_0       &   =         &  \{ {\bf X}^0 {\bf Z}^a        :  a = 1, 2, \ldots, p-1 \} 
    \nonumber \\                   
{\cal V}_1       &   =         &  \{ {\bf X}^a {\bf Z}^0        :  a = 1, 2, \ldots, p-1 \}   
    \nonumber \\
{\cal V}_2       &   =         &  \{ {\bf X}^a {\bf Z}^a        :  a = 1, 2, \ldots, p-1 \} 
    \nonumber \\
{\cal V}_3       &   =         &  \{ {\bf X}^a {\bf Z}^{2a}     :  a = 1, 2, \ldots, p-1 \} 
              \\
                 &\vdots  & 
    \nonumber \\
 {\cal V}_{p-1}  &   =         &  \{ {\bf X}^a {\bf Z}^{(p-2)a} :  a = 1, 2, \ldots, p-1 \} 
    \nonumber \\  
 {\cal V}_{p}    &   =         &  \{ {\bf X}^a {\bf Z}^{(p-1)a} :  a = 1, 2, \ldots, p-1 \} 
    \nonumber 
           \end{eqnarray} 
Each of the $p+1$ sets ${\cal V}_0, {\cal V}_1, \ldots, {\cal V}_{p}$ can be put in a one-to-one correspondance 
with one basis of the complete set of $p+1$ MUBs. In fact, ${\cal V}_0$ is associated with the computational basis 
$B_p$; furthermore, in view of 
           \begin{eqnarray}  
{\bf V_{0 a}} \in {\cal V}_{a + 1}, \quad a = 0, 1, \ldots, p-1 
           \end{eqnarray} 
it follows that ${\cal V}_1$, ${\cal V}_2$, $\ldots$, ${\cal V}_{p}$ are associated with the $p$ remaining MUBs 
$B_{00}$, $B_{01}$, $\ldots$, $B_{0p-1}$, respectively.}

Keeping into account the fact that the set $\{ {\bf X}^a {\bf Z}^b : a,b = 0, 1, \ldots, p-1 \} \setminus \{ {\bf X}^0 {\bf Z}^0 \}$
spans the Lie algebra of the special linear group $SL(p, \mathbb{C})$, we have the next theorem.

{\bf Theorem 4}. {\it For $d=p$ with $p$ a prime integer, the Lie algebra 
$sl(p, \mathbb{C})$ of the group $SL(p, \mathbb{C})$ can be decomposed into a sum 
(vector space sum indicated by $\uplus$) of $p+1$ abelian subalgebras each of dimension $p-1$, i.e., 
                  \begin{eqnarray}
sl(p, \mathbb{C}) \simeq 
{ v}_0     \uplus 
{ v}_1     \uplus 
\ldots     \uplus      
{ v}_{p}     
                  \end{eqnarray}
where the $p+1$ subalgebras ${ v}_0, { v}_1, \ldots, { v}_p$ are 
Cartan subalgebras generated respectively by the sets ${\cal V}_0, {\cal V}_1, \ldots, {\cal V}_{p}$ 
containing each $p - 1$ commuting matrices.}

The latter result can be extended when $d = p^e$ with $p$ a prime integer and 
$e$ an integer ($e \geq 2$): there exists a decomposition of $sl(p^e, \mathbb{C})$ into $p^e + 1$ 
abelian subalgebras of dimension $p^e - 1$ (cf.~\citep{PateraZassenhaus1988,Boykinetal2007,Kibler2009}). 

\section{Conclusion}

The quadratic discrete Fourier transform studied in this chapter can be considered 
as a two-parameter extension, with a quadratic term, of the usual discrete Fourier 
transform. In the case where the two parameters are taken to be equal to zero, the 
quadratic discrete Fourier transform is nothing but the usual discrete Fourier 
transform. The quantum quadratic discrete Fourier transform plays an important role 
in the field of quantum information. In particular, such a transformation in prime 
dimension can be used for obtaining a complete set of mutually unbiased bases. It 
is to be mentioned that the quantum quadratic discrete Fourier transform also arises 
in the determination of phase operators for the groups $SU(2)$ and $SU(1,1)$ in 
connection with the representations of a generalized oscillator algebra 
\citep{DaoKib2010,AKW2010}. As an open question, it should be worth 
investigating the relation between the quadratic discrete Fourier transform and the 
Fourier transform on a finite ring or a finite field.

\section*{Acknowledgements}

The author is greatly indebted to M. Daoud, M. Planat, O. Albouy, M. Saniga, N.M. Atakishiyev and K.B. Wolf 
for collaboration on various facets of the material contained in this chapter.

\section*{Appendix: Wigner-Racah algebra of $SU(2)$ in the $\{ j^2 , x \}$ scheme}

In this self-contained Appendix, the bar does not indicate complex conjugation. Here, complex conjugation 
is denoted with a star. 

The Wigner-Racah algebra of the group $SU(2)$ in the $SU(2) \supset U(1)$ or $\{ j^2 , j_z \}$ scheme
is well known. It corresponds to the use of bases of type $B_{2j+1}$ resulting from the simultaneous 
diagonalization of the Casimir operator $j^2$ and of the Cartan generator $j_z$ of $SU(2)$. Any change 
of basis of type
\begin{eqnarray}
| j , \mu \rangle = \sum_{m=-j}^{j} | j , m \rangle \langle j , m | j , \mu \rangle 
\label{transfo jm jmu}
\end{eqnarray}
(where for fixed $j$ the elements $\langle j , m | j , \mu \rangle$ define a $(2 j + 1)\times(2 j + 1)$ unitary 
matrix) leads to another acceptable scheme for the Wigner-Racah algebra of $SU(2)$. In this scheme, the matrices 
of the irreducible representation classes of $SU(2)$ take a new form as well as the coupling coefficients 
(and the associated $3-jm$ symbols). For instance, the Clebsch-Gordan or coupling coefficients 
$(j_1 j_2 m_1 m_2 | j m)$ are simply replaced by 
   \begin{eqnarray} 
(j_1 j_2 \mu_1 \mu_2 | j \mu) = 
\sum_{m_1=-j_1}^{j_1} 
\sum_{m_2=-j_2}^{j_2} 
\sum_{m  =-j  }^{j  }     (j_1 j_2 m_1 m_2 | j m) \nonumber \\
                   \langle j_1 , m_1 | j_1 , \mu_1 \rangle^* \,
                   \langle j_2 , m_2 | j_2 , \mu_2 \rangle^* \,
                   \langle j   , m   | j   , \mu   \rangle
   \end{eqnarray}
when passing from the $\left\{ j m   \right\}$ quantization to the 
                      $\left\{ j \mu \right\}$ quantization 
while the recoupling coefficients, and the corresponding $3(n-1)-j$ symbols, 
for the coupling of $n$ ($n \geq 3$) angular momenta remain invariant. The adaptation 
to the $\left\{ j \mu \right\}$ quantization scheme afforded by Eq.~(\ref{transfo jm jmu}) is
transferable to $SU(2)$ irreducible tensor operators. This yields the
Wigner-Eckart theorem in the $\{ j \mu \}$ scheme.  

We give here the basic ingredients for developing the Wigner-Racah algebra of $SU(2)$ in the $\{ j^2 , v_{00} \}$ 
or $\{ j^2 , x \}$ scheme. For such a scheme, the vector $| j , \mu \rangle$ is of the form
$| j \alpha ; 00 \rangle$ so that the label $\mu$ can be identified with $\alpha$. Thus, the 
inter-basis expansion coefficients $\langle j , m   | j , \mu \rangle$ are 
\begin{eqnarray}
\langle j , m | j \alpha ; 00 \rangle = {1 \over \sqrt{2 j + 1}} q^{(j+m) \alpha} 
= {1 \over \sqrt{2 j + 1}} 
\exp \left[ \frac{2 \pi {\rm i}}{2 j + 1} (j+m) \alpha \right]
\label{jm jalpha}
\end{eqnarray}
with 
$m      = j , j  - 1, \ldots,        -j$ 
and 
$\alpha = 0, 1, \ldots, 2j$. Equation (\ref{jm jalpha}) corresponds to the unitary 
transformation (\ref{j alpha r a in terms of jm}) with $r=a=0$, that allows to pass from the standard basis 
$B_{2j+1}$ to the nonstandard basis $B_{00}$. Then, the Clebsch-Gordan coefficients in the $\{ j^2, v_{00} \}$ 
scheme are 
  \begin{eqnarray}
\left( j_1 j_2 \alpha_1 \alpha_2 |j_3 \alpha_3 \right) = 
  {1 \over \sqrt{(2j_1 + 1) 
                 (2j_2 + 1) 
		             (2j_3 + 1)}} 
	\sum_{m_1=-j_1}^{j_1} 
  \sum_{m_2=-j_2}^{j_2} 
  \sum_{m_3=-j_3}^{j_3} 
		                               \nonumber          \\        
						(q_1)^{- (j_1 + m_1) \alpha_1}      \>
            (q_2)^{- (j_2 + m_2) \alpha_2}      \>
	          (q_3)^{  (j_3 + m_3) \alpha_3}      \>
   ( j_1 j_2 m_1 m_2 | j_3 m_3 ) 
  \end{eqnarray}
where the various $q_k$ are given in terms of $j_k$ by
       \begin{eqnarray}
q_k = \exp \left( {2 \pi {\rm i} \over 2 j_k + 1 } \right), \quad k = 1,2,3
       \end{eqnarray} 
The symmetry properties of the coupling coefficients $( j_1 j_2 \alpha_1 \alpha_2 | j_3 \alpha_3 )$
cannot be expressed in a simple way (except the symmetry under the interchange 
$j_1 \alpha_1 \leftrightarrow j_2 \alpha_2$). Therefore, it is interesting to introduce the 
following $\overline{f}$ symbol through 
  \begin{eqnarray}
  \overline{f} 
  \begin{pmatrix}
  j_1     &j_2     &j_3     \cr
  \alpha_1&\alpha_2&\alpha_3\cr
  \end{pmatrix} = 
  {1 \over {\sqrt{(2j_1 + 1) (2j_2 + 1) (2j_3 + 1)} } } 
  \sum_{m_1 = -j_1}^{j_1} 
  \sum_{m_2 = -j_2}^{j_2} 
  \sum_{m_3 = -j_3}^{j_3}  
  \nonumber \\  
         (q_1)^{ - (j_1 + m_1)\alpha_1 } \>
         (q_2)^{ - (j_2 + m_2)\alpha_2 } \>
         (q_3)^{ - (j_3 + m_3)\alpha_3 } 
  \begin{pmatrix}
  j_1&j_2&j_3\cr
  m_1&m_2&m_3\cr
  \end{pmatrix}       
  \label{def of fbar}
  \end{eqnarray}
where the $3-jm$ symbol on the right-hand side of (\ref{def of fbar}) is an ordinary 
Wigner symbol for the $SU(2)$ group in the $\{ j^2, j_z \}$ scheme. (The $\overline{f}$ symbol 
is to the $\{ j^2, x \}$ scheme what the $\overline{V}$ symbol of Racah is to the 
$\{ j^2, j_z \}$ scheme.) The $\overline{f}$ symbol exhibits the same symmetry properties under 
permutations of its columns as the $3-jm$ Wigner symbol (identical to the $\overline{V}$ Racah symbol): 
Its value is multiplied by $(-1)^{j_1 + j_2 + j_3}$ under an odd permutation 
and does not change under an even permutation. In contrast to the $3-jm$ symbol, not all the values of the 
$\overline{f}$ symbol are real. In this respect, the $\overline{f}$ symbol behaves under complex conjugation 
as 
  \begin{eqnarray}
  \overline{f} 
  \begin{pmatrix}
  j_1     &j_2     &j_3     \cr
  \alpha_1&\alpha_2&\alpha_3\cr
  \end{pmatrix}^* =  (-1)^{j_1 + j_2 + j_3}  \> (q_1)^{\alpha_1} (q_2)^{\alpha_2} (q_3)^{\alpha_3} \>
  \overline{f} 
  \begin{pmatrix}
  j_1     &j_2     &j_3     \cr
  \alpha_1&\alpha_2&\alpha_3\cr
  \end{pmatrix}
  \end{eqnarray}
Other properties (e.g., orthogonality properties, connection with the Clebsch-Gordan coefficients and 
the Herring-Wigner tensor, etc.) of the $\overline{f}$ symbol and its relations with $3(n-1)-j$ symbols 
for $n \geq 3$ can be derived along the lines developed in \citep{Kibler1968}.

\end{document}